\documentclass[fleqn,usenatbib]{mnras}
\usepackage{newtxtext,newtxmath}
\usepackage[T1]{fontenc}

\usepackage{graphicx}
\usepackage{amsmath}
\usepackage{subfigure}
\usepackage{enumitem}
\newcommand{\orb}{${\Omega}$\,} 
\newcommand{\spin}{${\omega}$\,} 
\newcommand{\twosp}{${2\omega}$\,} 
\newcommand{\be}{${\omega-\Omega}$\,} 
\newcommand{\twobe}{${2(\omega-\Omega)}$\,} 
\newcommand{\porb}{P$_{\Omega}$\,} 
\newcommand{\psp}{P$_{\omega}$\,} 
\newcommand{\pbe}{P$_{\omega-\Omega}$\,} 
\usepackage{stfloats} 

\title[V515 And: An IP in the period gap]{V515 And: An Intermediate Polar in the Period Gap Exhibiting Outbursts}

\author[Rao et al.]{
Srinivas M Rao,$^{1,2}$\thanks{E-mail: srinivas@aries.res.in, srinivas22546@gmail.com},
Jeewan C Pandey$^{1}$\thanks{E-mail: jeewan@aries.res.in},
Nikita Rawat$^{3}$,
Simone Scaringi$^{4}$,
Arti Joshi$^{5}$,
David A.\,H. Buckley$^{3}$,
\newauthor
and Ajay Kumar Singh$^{6}$
\\
$^{1}$Aryabhatta Research Institute of Observational Sciences(ARIES), Nainital 263001, India\\
$^{2}$Mahatma Jyotiba Phule Rohilkhand University, Bareilly 243006, India\\
$^{3}$South African Astronomical Observatory, PO Box 9, Observatory, Cape Town, 7935, South Africa\\
$^{4}$Centre for Extragalactic Astronomy, Department of Physics, Durham University, South Road, Durham DH1 3LE, UK\\
$^{5}$Institute of Astrophysics, Pontificia Universidad Cat\'olica de Chile, Av. Vicu\~na Mackenna 4860, 7820436 Macul, Santiago, Chile\\
$^{6}$Department of Applied Physics/Physics, Bareilly College, Bareilly-243001, India
}

\date{Accepted XXX. Received YYY; in original form ZZZ}

\pubyear{\the\year{}}

\begin{document}
\label{firstpage}
\pagerange{\pageref{firstpage}--\pageref{lastpage}}
\maketitle

\begin{abstract}
Using long-term observations from the Transiting Exoplanet Survey Satellite (TESS) along with spectroscopic observations from the 3.6 m Devasthal Optical Telescope (DOT), we present a comprehensive time-series and spectral analysis of the intermediate polar V515 And. Our analysis reveals that V515 And resides within the period gap, with the detection of its orbital period of 2.73116 h. Additionally, we confirm the earlier findings of the spin and beat periods to be 465.4721 s and 488.6067 s, respectively.  The time-resolved timing analysis reveals that V515 And undergoes changes in its accretion geometry, not only between different TESS sectors but also within individual sector observations.  The system exhibits a transition in the dominant accretion mode, switching between disc-fed and stream-fed accretion. In the TESS light curve, we identify two successive outburst-like episodes, each persisting for roughly a day and reaching peak luminosities of $2.7\times10^{33}$ and $1.9\times10^{33}$ erg s$^{-1}$. Our analysis suggests that these bursts belong to the recently proposed class of micronovae. The optical spectrum of V515 And is characterised by strong Balmer and He II emission lines and shows an inverse Balmer decrement indicating the magnetic nature of the source. 

\end{abstract}




\section{Introduction}\label{sec:intro}
Intermediate polars (IPs) are a subclass of the magnetic cataclysmic variable (MCVs) where the magnetic field strength of the primary component of the binary, white dwarf (WD), ranges from $\sim$0.1 to 10 megagauss (MG) and accretes mass from the secondary red dwarf \citep{Patterson1994}. Due to the influence of the magnetic field of the WD, the accreted material accumulates near the magnetic poles. The accretion regime is either via disc, stream, or both. The disc-fed accretion occurs via a Keplerian disc, and this disc is disrupted at a specific point known as the magnetospheric radius of WD, and from this point, the material follows the magnetic field lines of the WD \citep{Hellier1989a}. In contrast, in the stream-fed accretion, the material is directly accreted onto the magnetic poles of the WD following the magnetic field lines \citep{Rosen1988}. In stream-fed accretion, if the material skims over the disc and accretes through the disc as well, then the accretion is called disc overflow \citep{Hellier1989b}. The power at the spin (\spin) and the beat (\be) frequencies in the power spectra help us identify the probable accretion geometry of the system \citep{Hellier1991, Hellier1993}. The magnetically channelled accretion column free-falls with supersonic velocities close to the surface of the WD, which leads to the formation of shocks. X-rays are emitted in the post-shock region as the matter cools via thermal bremsstrahlung and cyclotron radiation \citep{Aizu1973, Wu1994, Cropper1999}. The X-rays may be reprocessed from other regions of the WD, such as the WD itself or the accretion disc surrounding it, resulting in optical emission.

\begin{table*}
    \centering
    \caption{Log of observations of V515 And using TESS.}
    \begin{tabular}{ c c c c c}
    \hline
      Sector   & Start Date of Obs. & Start Time of Obs. & End Date of Observation & End Time of Observation  \\
                & (YYYY-MM-DD) & (UT) & (YYYY-MM-DD) & (UT) \\  
    \hline
     17 & 2019-10-08 & 04:29:46 & 2019-11-02 & 04:22:01 \\
     57* & 2022-09-30 & 20:34:42 & 2022-10-29 & 14:51:17 \\
     84 & 2024-10-01 & 02:13:36 & 2024-10-26 & 20:16:09 \\
     85 & 2024-10-27 & 01:24:09 & 2024-11-21 & 13:09:25 \\                  
    \hline
    \end{tabular}
    ~~\\
    \text{*} The data is available for both 120 and 20 s cadence.
    \label{tab:obslog}
\end{table*}

\begin{figure*}
    \centering
    \subfigure[TESS light curve]{\includegraphics[width=\linewidth]{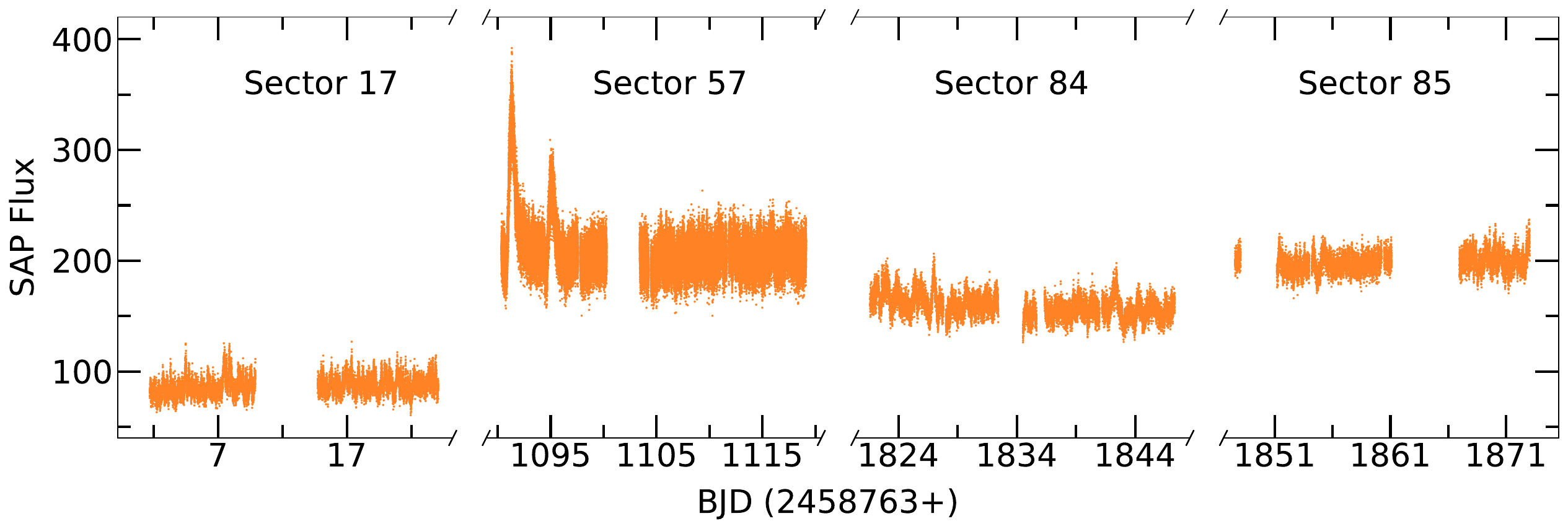}\label{fig: tess_lc_V515}}
    \subfigure[Outburst]{\includegraphics[width=\linewidth]{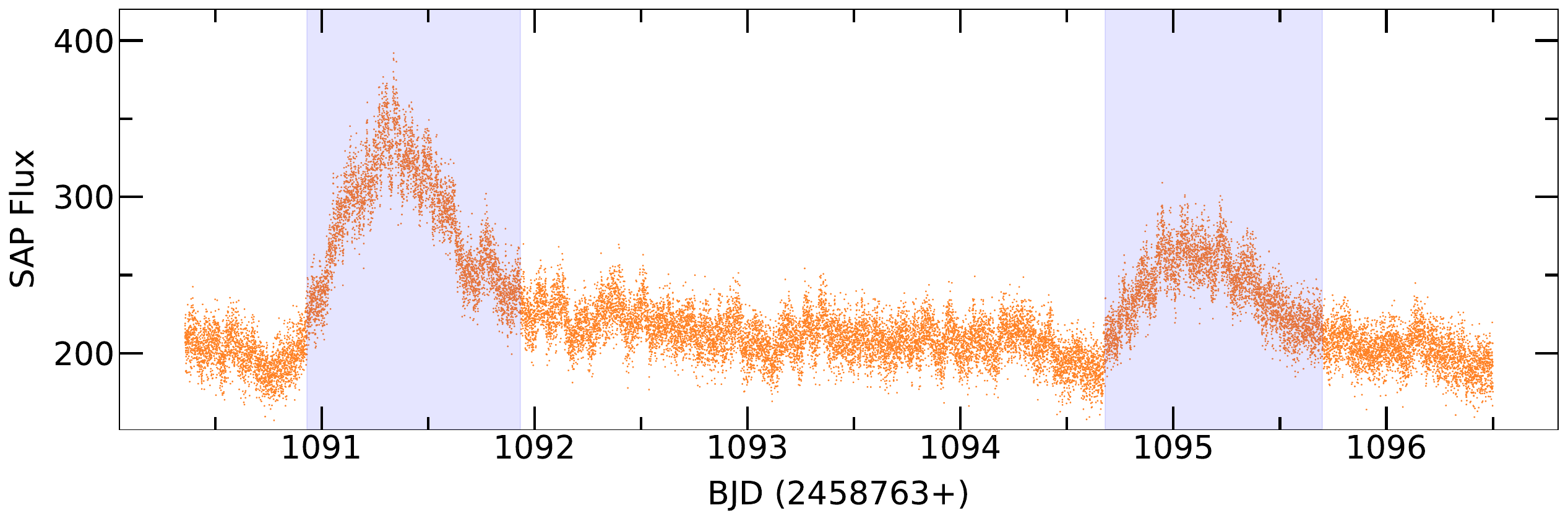}\label{fig: lc_V515_burst}}
    \caption{(a) TESS light curve of all sectors. (b) Zoomed-in section from sector 57 showing the two outbursts in shaded regions.}
    \label{fig:lc_V515}
\end{figure*}

The orbital period (\porb) of IPs generally ranges from 1.35-48 h \footnote{\url{https://asd.gsfc.nasa.gov/Koji.Mukai/iphome/catalog/alpha.html}}. There is a significant dearth of sources in the 2-3 h range, and this is known as the period gap \citep{Warner1991}. However, the IP Paloma with a spin-to-orbital period ratio of 0.83 was the only earlier known source in the period gap \citep{Schwarz2007,Joshi2016}. This period gap is also observed for non-magnetic cataclysmic variables (NMCVs), but it is less obvious in the case of polars \citep{Schreiber2024}. Recent studies by \cite{Schreiber2024} have refined the boundaries of the period gap in CVs to 2.45-3.18 hrs. The onset of the period gap is explained by the disruptive magnetic braking (DMB) model, which indicates a significant decrease or complete cessation of mass transfer from the secondary. Gravitational radiation brings the stars closer together, allowing mass transfer to resume at a much lower rate \citep{Kolb1993}. 

In this paper, we investigate the accretion properties of V515 And (= XSS J00564+4548), a system that appears to reside within the period gap.
It was identified in the RXTE all-sky survey with the spin period (\psp) of the WD as $\sim$480 s \citep{Bikmaev2006}. \cite{Butters2008} found \psp to be 465.68 $\pm$ 0.07 s and also obtained a probable beat period (\pbe) of 489.0 $\pm$ 0.7 s. However, \cite{Bonnet-Bidaud2009} found a different \psp in X-rays of 469.75 $\pm$ 0.26 s. Based on approximately 1.3 years of white light observations separated by a one-year gap, \cite{Kozhevnikov2012} reported refined values for the \psp and \pbe of V515 And as 465.48493 $\pm$ 0.00007 s and 488.61822 $\pm$ 0.00009 s, respectively. Using these results, the orbital period (\porb) of V515 And is estimated to be between 2.62 and 2.73 h, which places it in the period gap.  This finding necessitates further investigation of V515 And using high-cadence, long-baseline TESS data.  The presence of two outburst-like features in the TESS light curve further motivated us to do this study.

We organise the paper as follows: in section \ref{sec:obs}, we describe the observations and data. Section \ref{sec:ana} contains the analysis and results. Finally, we present the discussion and conclusions in sections. \ref{sec:dis} and \ref{sec:conc}, respectively.

\begin{figure*}
    \centering
    \includegraphics[width=\textwidth]{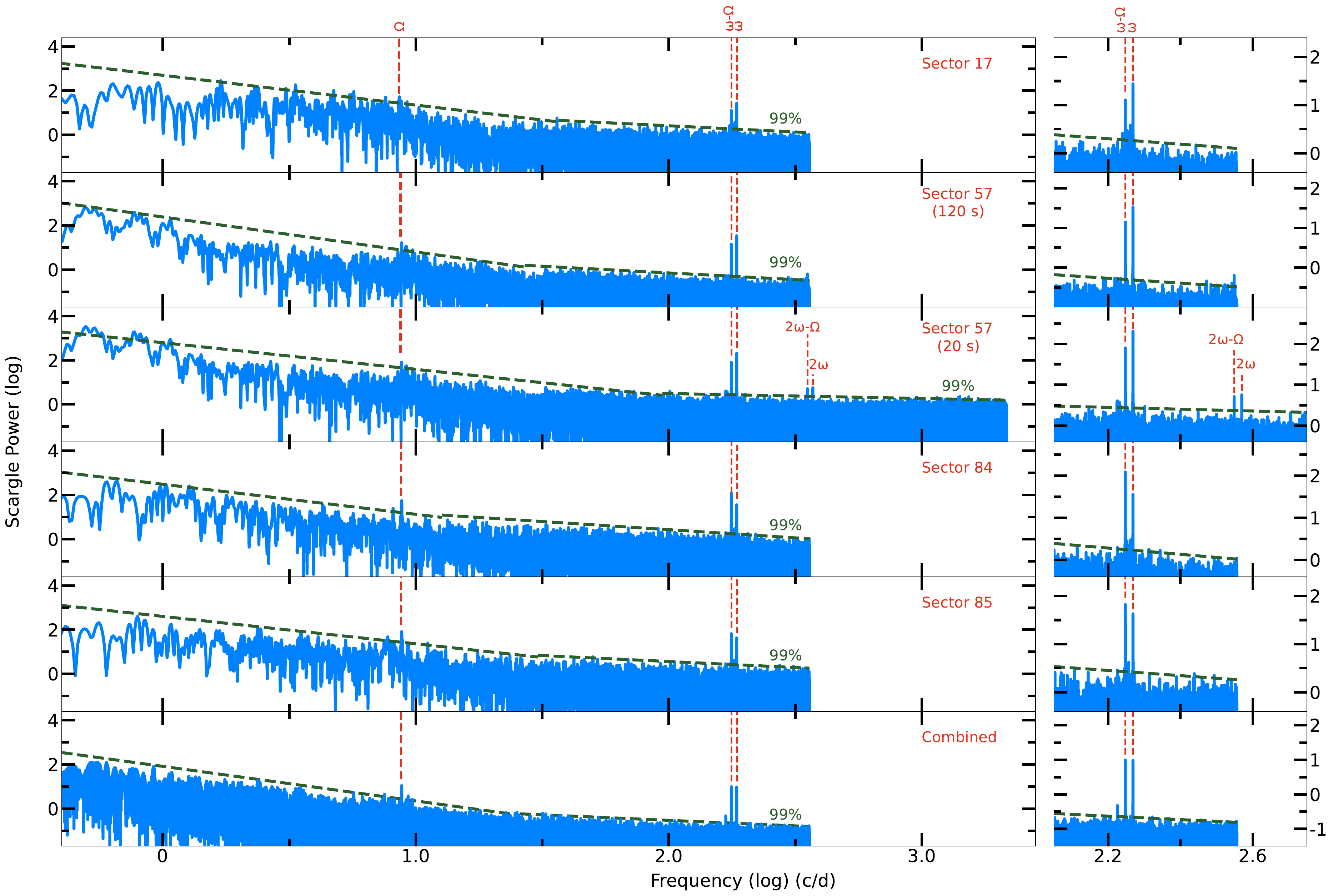}
    \caption{TESS power spectra of different sectors and the combined dataset. The identified frequencies are marked with red vertical dashed lines. The green dashed line represents the 99\% confidence level. The right panel shows the zoomed-in part near the  \spin and \be frequencies. }
    \label{fig:ps_V515} 
\end{figure*}

\section{Observations and Data}\label{sec:obs}
\subsection{Timing observations}
V515 And was observed by TESS for four sectors. The log of observations is given in Table \ref{tab:obslog}.
The TESS \citep{Ricker2015} instrument consists of four wide-field CCD cameras that can image a region of the sky, measuring $24^\circ \times 96^\circ$. TESS observations are broken up into sectors, each lasting two orbits, or about 27.4 days and conduct their downlink of data while at perigee. This results in a small gap in the data compared to the overall run length. The data is stored in the Mikulski Archive for Space Telescopes data\footnote{\url{https://mast.stsci.edu/portal/Mashup/Clients/Mast/Portal.html}} with unique identification number `TIC 196278926'. TESS provides data in two forms, namely Simple Aperture Photometry (SAP) flux and the pipeline-processed Pre-search Data Conditioned Simple Aperture Photometry (PDCSAP) flux. PDCSAP flux is the SAP flux values corrected for instrumental variations\footnote{See section 2.1 of the TESS archive manual at \url{https://outerspace.stsci.edu/display/TESS/2.1+Levels+of+data+processing}}. While performing corrections in the SAP flux values, the pipeline can sometimes also remove systematic trends \citep{Jenkins2016}. Hence, for our analysis, we have used SAP flux, which had quality flags marked as `0'. 
 

\subsection{Spectroscopic observations}
Spectroscopic observations of V15 And were taken using the ARIES-Devasthal Faint Object Spectrograph and Camera (ADFOSC; \citealt{Omar2019}), mounted on the 3.6 m Devasthal optical telescope (DOT; \citealt{Kumar2018}). It is comprised of a 4k $\times$4k deep-depletion CCD camera providing a 0.2 arcseconds/pixel scale with a 2 $\times$2 binning \citep{Panchal2023}. The observations were done with a slit 1.5" wide and 8" long and a 132R-600 gr/mm grism which covers a wavelength range of 3500-7000 \AA, centred at 4880 \AA. A total of three spectra were obtained on three consecutive nights - 25, 26 and 27th January 2025, each having an exposure of 40 min. These data were obtained $\sim$4 months after the last TESS Sector 85 observation. The bias and flat frames were taken for pre-processing, and the HgAr lamp was used for wavelength calibration. For flux calibration, a standard star (HZ 2) was observed on the same night, using the same slit and grism configurations as those for the source. The reduction was done using Image Reduction and Analysis Facility (IRAF) software (\citealt{Tody1986, Tody1993}).


\begin{table*}
    \centering
    \caption{Periods obtained corresponding to the dominant peaks of the LS power spectra of TESS.}
    \label{tab:ps_V515}
    \begin{tabular}{c c c c c c c c c}
    \hline
    Sector & Cadence (s) & $P_\Omega$(h) & $P_\omega$(s) & $P_{\omega-\Omega}$(s) & $P_{2\omega}$(s) & $P_{2(\omega-\Omega)}$(s) \\
    \hline
    17 & 120 & $2.740 \pm 0.003$ & $465.46 \pm 0.03$ & $488.60 \pm 0.03$ & - & - \\
    57 & 120 & $2.731 \pm 0.002$ & $465.46 \pm 0.02$ & $488.59 \pm 0.02$ & - & - \\
    57 & 20  & $2.731 \pm 0.003$ & $465.46 \pm 0.02$ & $488.59 \pm 0.02$ & $232.735 \pm 0.005$ & $244.296 \pm 0.006$\\
    84 & 120 & $2.728 \pm 0.003$ & $465.47 \pm 0.02$ & $488.62 \pm 0.03$ & - & - \\ 
    85 & 120 & $2.731 \pm 0.003$ & $465.46 \pm 0.02$ & $488.56 \pm 0.02$ & - & - \\
  Mean$^*$ &     & $2.731 \pm 0.001$ & $465.46 \pm 0.01$ & $488.59 \pm 0.01$ & - & - \\
    Combined$^\dagger$ & & $2.73116 \pm 0.00004$* & $465.4721 \pm 0.0003$ & $488.6067 \pm 0.0004$ & - & - \\
    \hline
    \end{tabular}\\
{$^*$ Weighted mean and corresponding weighted error of periods derived from different sectors. \\ 
$^\dagger$ Represents the periods derived from the power spectra of combined TESS observations of all sectors.\\} 
\end{table*}

\begin{figure*}
    \centering
    \subfigure[Sector 17]{\includegraphics[width=0.49\textwidth]{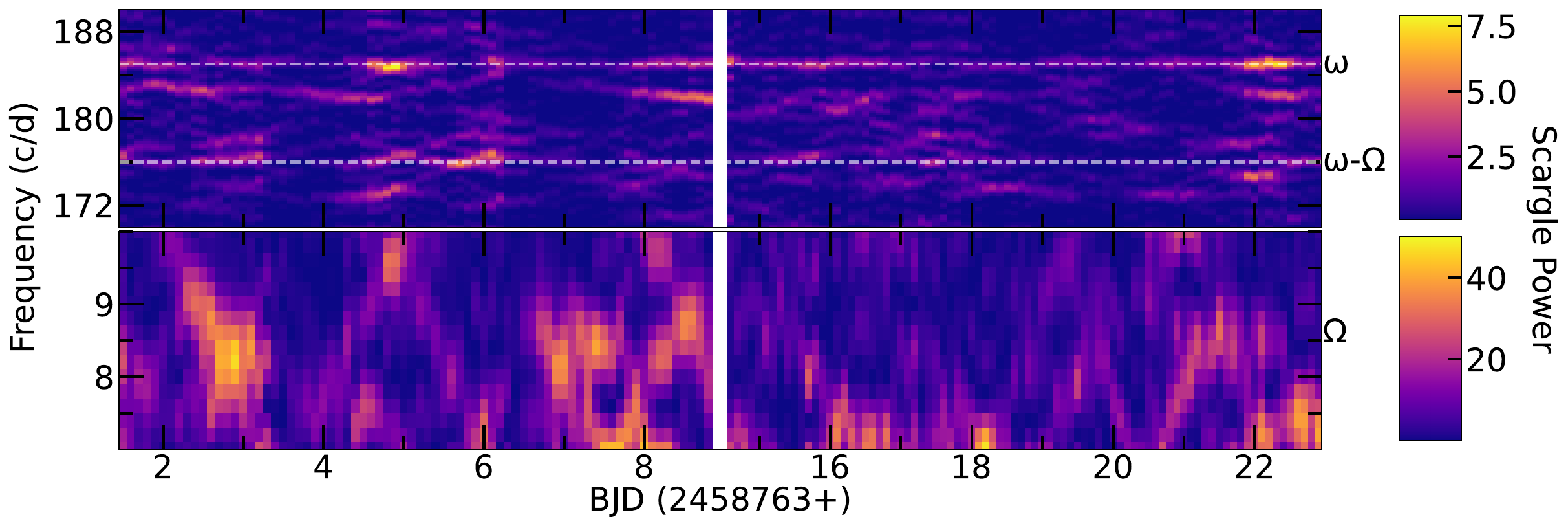}\label{fig:tps_17}}
    \subfigure[Sector 57]{\includegraphics[width=0.49\textwidth]{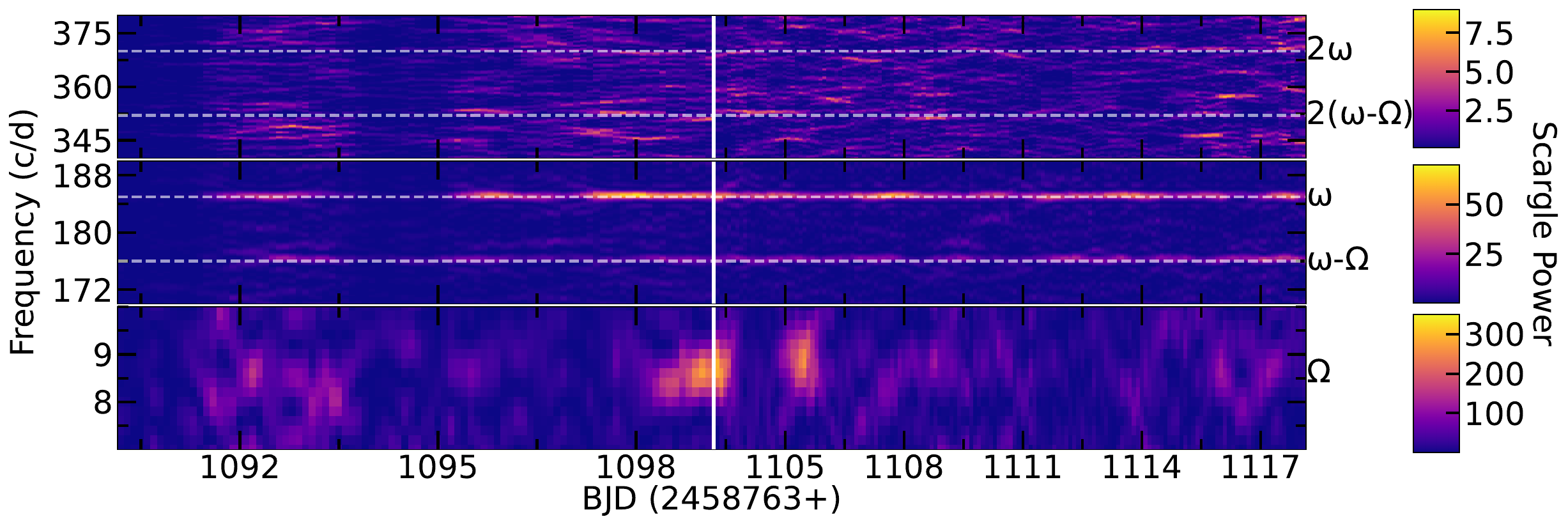}\label{fig:tps_57}}
    \subfigure[Sector 84]{\includegraphics[width=0.49\textwidth]{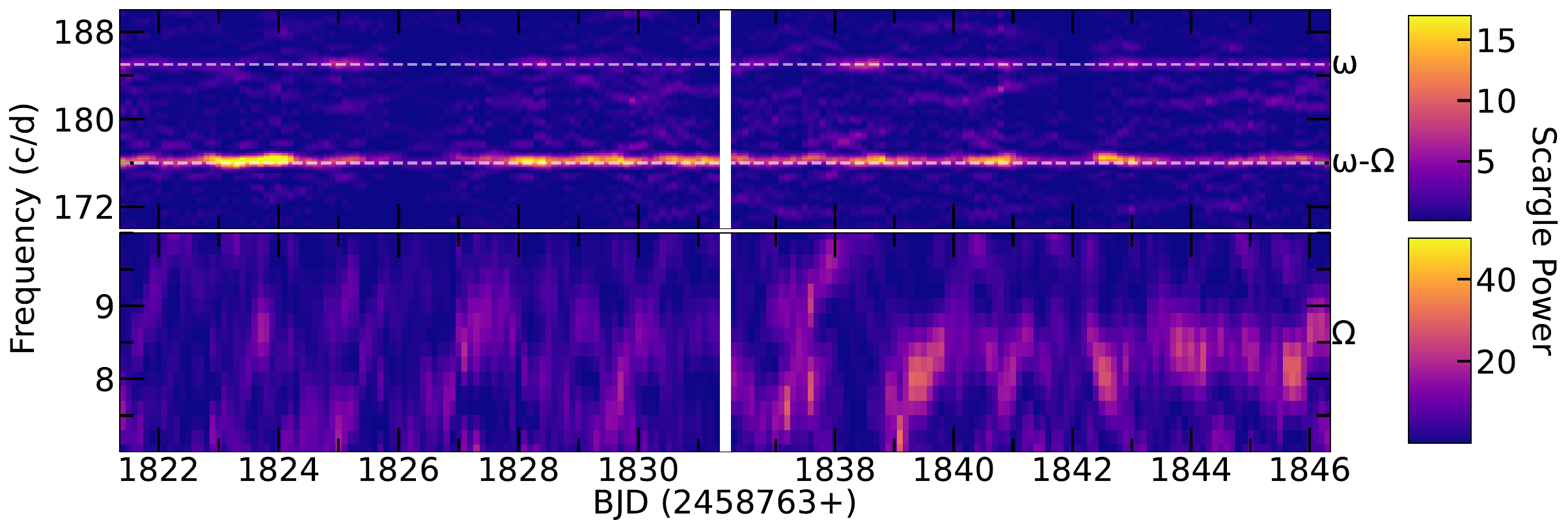}\label{fig:tps_84}}
    \subfigure[Sector 85]{\includegraphics[width=0.49\textwidth]{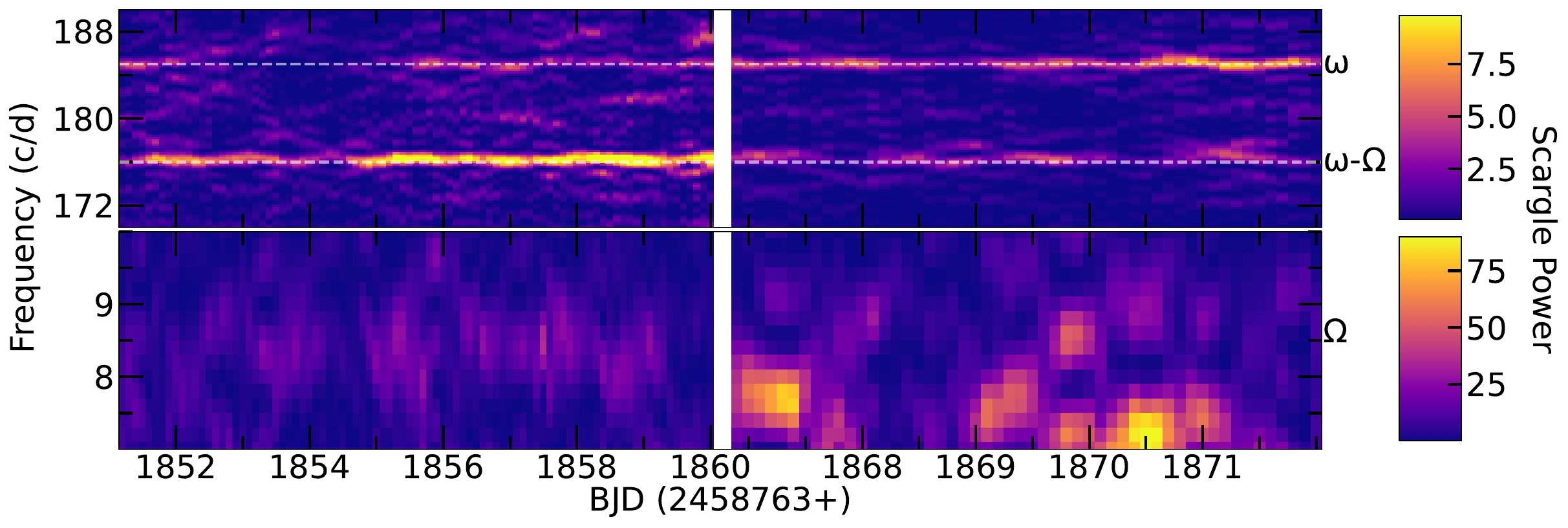}\label{fig:tps_85}}
    \caption{ Trailing power spectra from TESS, binned using a 1-day moving window with a step size of 0.1 days. The \spin, and \be frequencies, along with their identified harmonics, are shown with dashed lines.}
    \label{fig:tps_V515}
\end{figure*}

\section{Analysis and Results}\label{sec:ana}
\subsection{\textit{Timing analysis}}
\subsubsection{Light curve and power spectra}\label{V515_lc}
The combined light curve of V515 And from all sectors' observation is shown in \ref{fig: tess_lc_V515}. The variability in the light curve is clearly visible. Two burst-like features in the light curve of sector 57,  lasting $\sim$1 day, were noticed and shown in \ref{fig: lc_V515_burst}. To determine the periodicities of the underlying variabilities, the Lomb-Scargle (LS) periodogram method \citep{Lomb1976, Scargle1982} was applied to the light curves of the individual sectors and the combined dataset from all sectors. For sector 57,  the burst duration in the light curve was removed for the periodogram analysis.  The significance of peaks within the power spectra was determined using the methodology described by \cite{Vaughan2005}, assuming that TESS power spectra possess different types of noise \citep{Kalman2025}. We have only considered those peaks which were above the 99$\%$ significance level. The power spectra for all the sectors are shown in Figure \ref{fig:ps_V515}.

We primarily identified three significant frequencies at 8.87, 176.84, and 185.62 cycles/day (c/d) in all sectors as well as the combined dataset, which are designated as orbital (\orb), \be, and \spin, respectively. Recently, \cite{Bruch2025} also obtained these frequencies in his analysis. In the combined dataset, the \spin and \be frequencies have almost equal power. However, in sectors 17 and 57, the \spin and in sectors 84 and 85, the \be were found to be dominant frequencies in the power spectra. We identified that the \orb\ frequency of the system is consistent with the inferred \orb\ frequency using the \spin and \be\ frequencies. In addition to the above-mentioned frequencies, the power spectral analysis of the 20 s cadence data of sector 57 also shows the presence of harmonics of \spin\ and \be\ frequencies.  A few other frequencies towards the bluer end of the power spectra were found above the significance level, but none of these frequencies were consistent among the power spectra of all sectors. Furthermore, none of these frequencies were present in the combined power spectra. Therefore, we have not considered them for any further analysis. The periods corresponding to these frequencies are given in Table \ref{tab:ps_V515}.  We have also calculated the weighted mean and corresponding weighted error for all periods across different sectors, which are consistent with those derived from the combined data set.   


\begin{figure*}
    \centering
    \subfigure[Spin phase folded]{\includegraphics[width=0.49\textwidth]{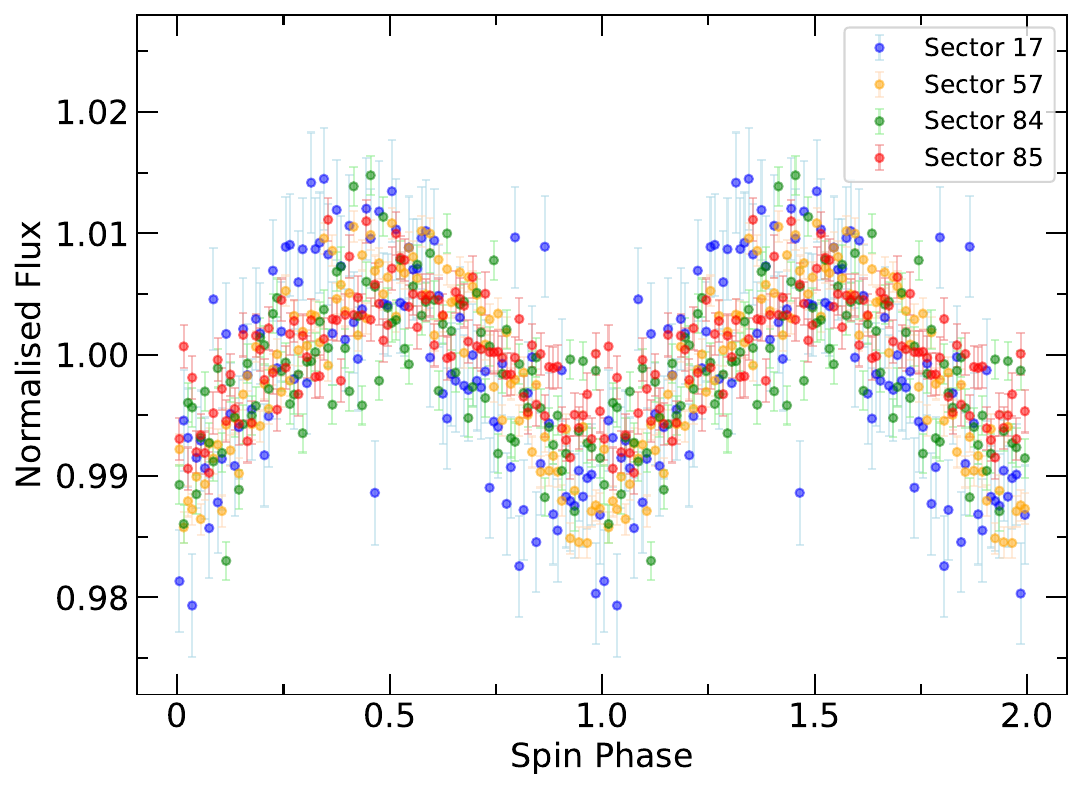}\label{fig:sp}}
    \subfigure[Beat phase folded]{\includegraphics[width=0.49\textwidth]{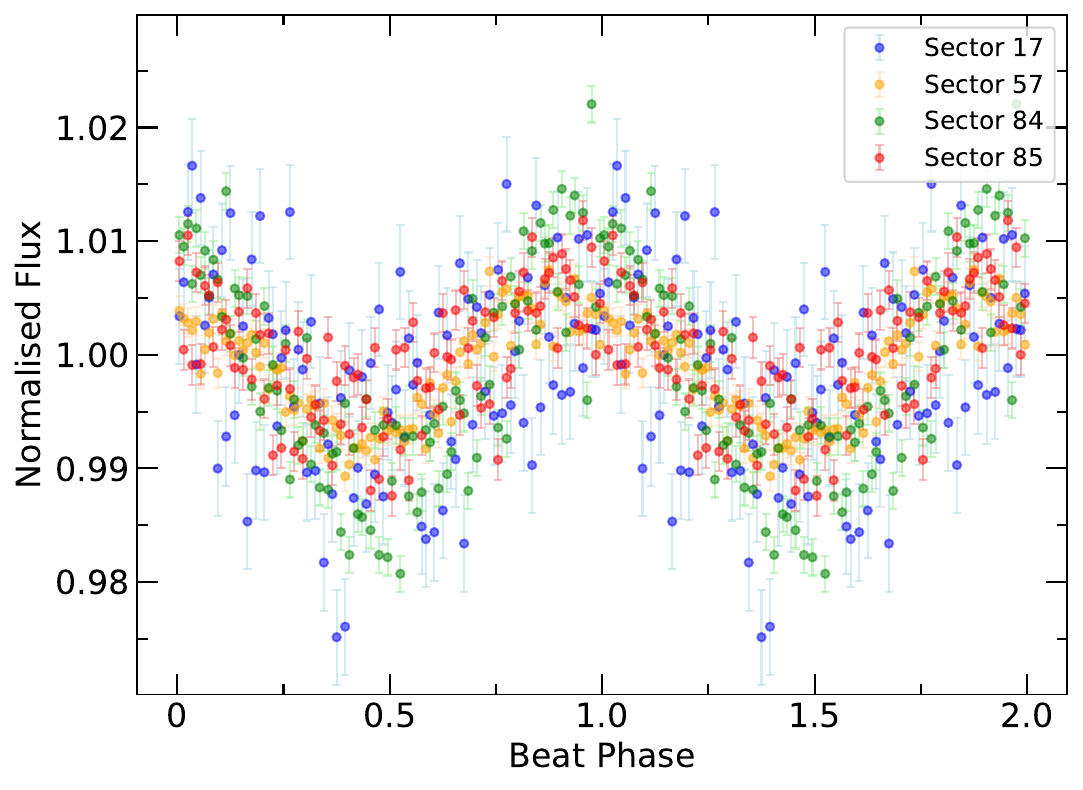}\label{fig:be}}
    \caption{ Spin and beat phase folded light curves of V515 And for each observed sector.}
    \label{fig:phase_sb}
\end{figure*}

\subsubsection{Trailed power spectra}\label{sec:V515_tps}
As the dominance of \spin and \be\ frequencies varied across power spectra of different sectors' data, a trailing power spectrum was calculated to study their evolution. Figure \ref{fig:tps_V515} shows the trailed power spectra of all four sectors of TESS observations. To obtain the trailed power spectra, we chose a window size of 1 day and moved it in increments of 0.1 day. We have shown only the frequency regions around which we obtained the \orb, \spin, and \be\ frequencies in the sector-wise power spectra. For sector 57, we show results for a 20 s cadence (see Figure \ref{fig:tps_57}), which also includes frequencies at \twosp\ and \twobe. In each sector, the \orb\ frequency was detected only in a few days. In sectors 17 and 57, the \spin\ frequency was more dominant than the \be\ frequency on most days; however, in sector 84, the \be\ frequency was more dominant on most days. The results for the 120 s cadence dataset in sector 57 were similar to those for the 20 s cadence dataset. Following sector 84, in the first half of sector 85, \be\ frequency remained dominant, whereas in the 2nd half, the dominance again switched to \spin\ frequency. During the period of bursts, there was the absence of both \spin and \be\ frequencies. The power of \spin and \be\ frequencies gradually decreases before the onset of the second burst, and then the power of both frequencies gradually increases after the end of the outburst.
 
\begin{figure*}
    \centering
    \subfigure[Spin phase folded]{\includegraphics[width=\textwidth]{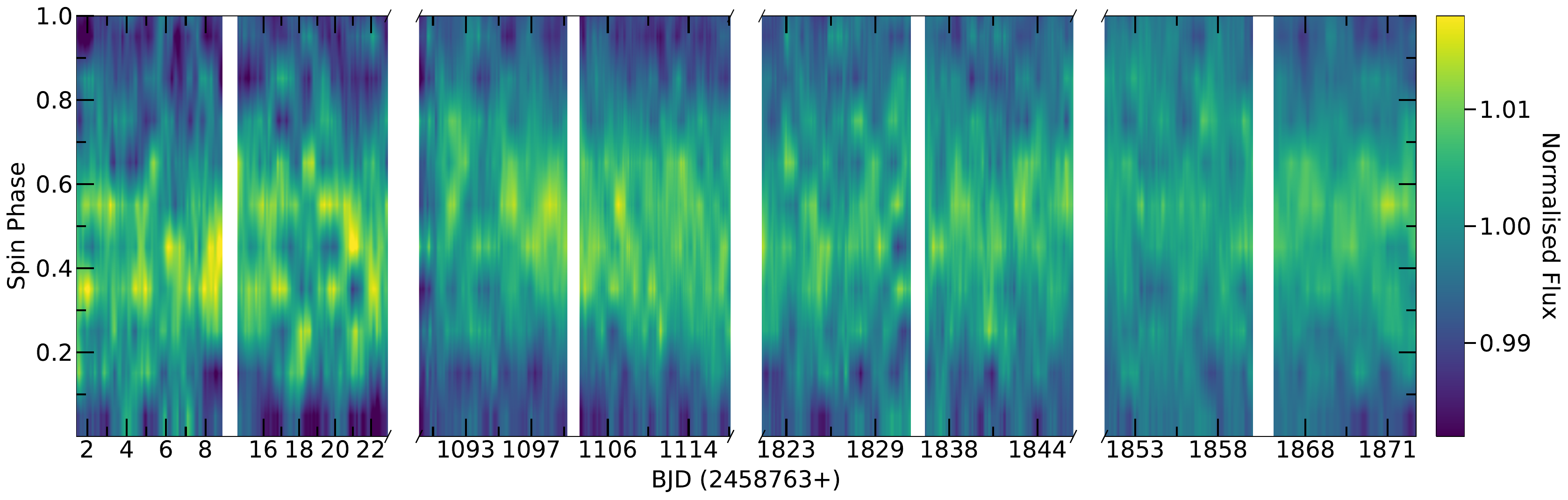}\label{fig:spfold}}
    \subfigure[Beat phase folded]{\includegraphics[width=\textwidth]{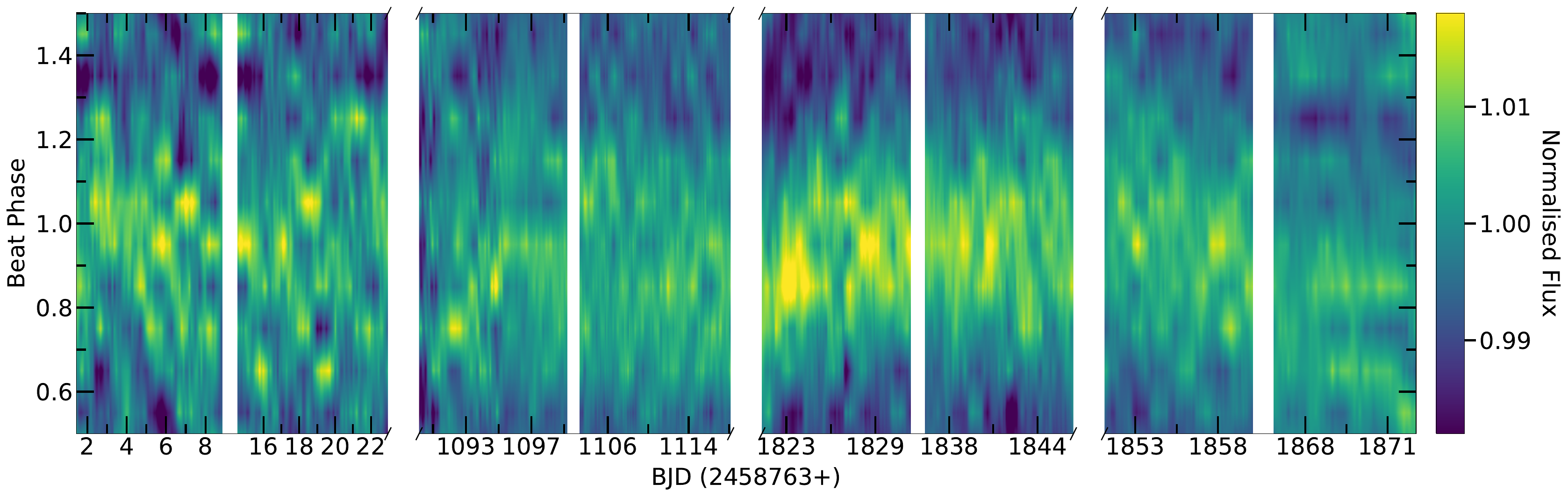}\label{fig:befold}}
    \subfigure[Pulse Fraction]{\includegraphics[width=\textwidth]{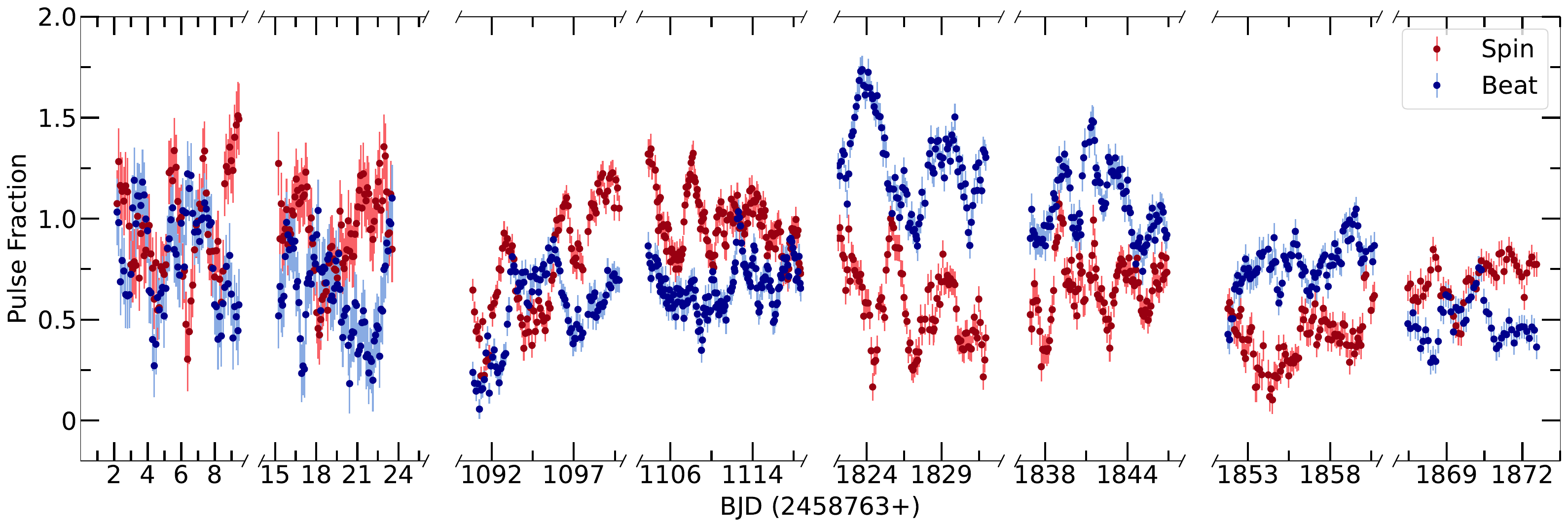}\label{fig:pf}}
    \caption{(a) and (b) Trailed spin and beat phase folded light curve of the entire dataset. (c) Variation of spin and beat pulse fraction. }
    \label{fig:phase_pf}
\end{figure*}

\subsubsection{Phase-folded light curves and pulse fraction}
Light curves were folded over \psp and \pbe, which were obtained from the combined dataset using the ephemeris given by \citep{Kozhevnikov2012}. The spin- and beat-phase-folded light curves for all sectors are shown in Figures \ref{fig:sp} and \ref{fig:be}, respectively. To make the trailed phase-folded light curve, we have followed a similar methodology to that discussed in Section \ref{sec:V515_tps} for obtaining trailed power spectra. The trailed spin and beat phase folded light curves of all the sectors are shown in Figure \ref{fig:spfold} and \ref{fig:befold}, respectively. A single peak pattern was observed in both phase-folded light curves, which was centred around 0.5 phase in spin-phased light curves and $\sim$ 0.9 phase in beat-phased light curve. For each phased light curve, we have derived the pulse fraction (PF) as

\begin{equation}
    PF = \frac{A_{max} - A_{min}}{A_{max} + A_{min}} \times 100 \% \label{eq:pf}
\end{equation}

Here, $A_{max}$ and $A_{min}$ are the maximum and minimum flux of the day-wise phased light curve, which were obtained by fitting a sinusoid on that phased light curve. The spin and beat pulse fraction for all the sectors is shown in Figure \ref{fig:pf}. In sector 17, the spin and beat pulse fractions were comparable, but on a few days, the spin pulse fraction was higher. In contrast, in the latter sectors, a distinction is observed between the values of spin and beat pulse fractions. On the majority of days in sector 57, the spin pulse fraction was dominant, whereas in sector 84, the beat pulse fraction was dominant. During the first half of sector 85, the beat pulse fraction was dominant, and in the latter half, the dominance changed to the spin pulse fraction.

\subsubsection{Characterising the Outbursts}\label{sec:V515_burst_tess}
 Though the TESS excels in providing precise relative photometry, it is advantageous to utilise data from another observatory to obtain absolute photometry from TESS observations. To convert the TESS data into flux, we need to calibrate them using simultaneous ground-based observations, following a methodology similar to that applied by several authors in the past \citep[such as][]{Scaringi2022b, Ilkiewicz2024, Irving2024, Veresvarska2024, Veresvarska2025}. We have used the simultaneous ASAS-SN g-band data for calibration. The g-band is centred at 477 nm and has a width of 100 nm. As the outbursts were present in sector 57, the availability of 20 s cadence data for that sector allowed us to calibrate with ground-based photometry.  We have selected only those ASAS-SN g-band data points that were obtained within 20 s of the TESS observations. The calibration was done for each half of the sector separately. Then, we fit a linear relationship between the two bands, and a direct conversion can be established by the relation
\begin{equation}
    F_{\text{ASAS-SN g}} (Jy) =  A \times F_{\text{TESS}} (e^- s^{-1})   + C \label{eqn:cal}
\end{equation}
This converts the TESS count rates into the ASAS-SN g-band magnitude. From the \textit{Gaia} DR3 parallax \citep{Gaia2023}, the distance of the source is calculated as 958$\pm$24 pc. Using the distance and TESS's bandwidth (500 nm), we calculated the luminosity in the g-band. The luminosity obtained here is underestimated because we have not considered any bolometric correction when cross-calibrating ASAS-SN and TESS, and we have assumed a flat emission spectrum while converting from spectral flux density to luminosity.

We have used the Bayesian Block algorithm \citep{Scargle1998, Scargle2013} to identify the start and end times of the outburst. The start and end times of the first outburst were found to be  BJD 2459453.9321 and 2459854.933, respectively; whereas to those of the 2nd outburst were BJD 2459857.678 and 2459858.697, respectively. The durations of the first and second outbursts were found to be 1.001 and 1.019 days, respectively. Both outbursts have a fast rise and slow decline, and there is a dip before the start of both outbursts.    The first outburst was found to have a higher amplitude than the latter. The peak optical luminosity of the first outburst was found to be  $(2.7 \pm 0.5) \times 10^{33} $ erg s$^{-1}$ and that of the second outburst was found to be $(1.9 \pm 0.4) \times 10^{33} $ erg s$^{-1}$. 
We have estimated the baseline luminosity during the burst by using luminosity values from the non-outburst phase. A running mean was computed, and the gaps were interpolated using the spline function.
We then subtract this baseline luminosity from the calibrated light curve. By integrating this baseline-subtracted luminosity during the outburst, we have obtained the total energy of the bursts. The energy of the first outburst was determined to be $(6.0 \pm 0.1) \times 10^{37}$ erg, whereas that of the 2nd outburst was determined to be $(2.9\pm 0.1) \times 10^{37}$ erg. 

\begin{table*}
\centering
\caption{Identification, flux, EW, and FWHM for emission features of the observed spectra of V515 And at three different epochs.}
\label{tab:opt_spect} 
\begin{tabular}{l ccc ccc ccc}
    \hline
    Identification & \multicolumn{3}{c}{25th January} & \multicolumn{3}{c}{26th January} & \multicolumn{3}{c}{27th January} \\
    \cline{2-4} \cline{5-7} \cline{8-10} 
                   & Flux & -EW & FWHM & Flux & -EW & FWHM & Flux & -EW & FWHM \\
    \hline 
    H$\gamma$ (4340\,\AA) & 5.2 & 5.88 & 922 & 3.9 & 4.13 & 815 & 3.3 & 4.66 & 818\\
    C III/N III (4640/50\,\AA) & 1.8 & 3.0 & 1217 & 2.0 & 3.17 & 1210 & 1.2 & 2.29 & 1094\\
    He II (4686\,\AA)  & 2.3 & 4.01 & 842 & 2.6 & 3.99 & 923 & 2.2 & 4.19 & 1141\\
    H$\beta$ (4861\,\AA)  & 4.2 & 7.91 & 835 & 3.7 & 6.80 & 797 & 3.9 & 7.41 & 820\\
    He I (4922\,\AA)   & 0.5 & 1.03 & 657 & 0.4 & 0.69 & 487 & 0.4 & 0.82 & 503\\
    He I (5875\,\AA)   & 0.9 & 3.33 & 726 & 1.0 & 3.50 & 682 & 1.0 & 3.65 & 714\\
    H$\alpha$ (6563\,\AA) & 3.9 & 17.56 & 793 & 3.5 & 14.64 & 747 & 3.8 & 17.55 & 776\\
    He I (6678\,\AA)   & 0.5  &  2.71 & 697 & 0.6 & 2.9 & 874 & 0.6 & 2.94 & 722\\
    \hline 
\end{tabular}
\par\medskip 
\small 
Note. Flux, EW, and FWHM are in the unit of $10^{-14}$ erg cm$^{-2}$ s$^{-1}$, \AA, and km s$^{-1}$, respectively.
\end{table*}

\begin{figure}
    \centering
    \includegraphics[width=\columnwidth]{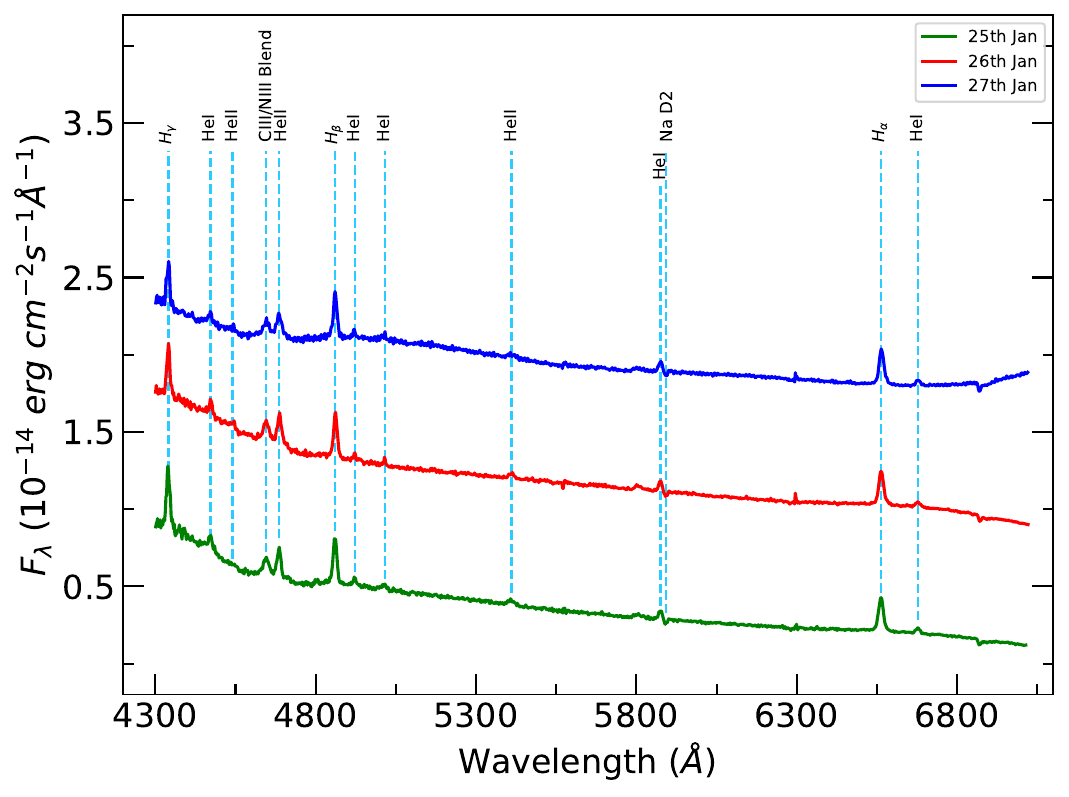}
    \caption{Optical spectra of V515 And obtained during three epochs. The colours represent different epochs of observations. Constants of 0.8 and 1.6 have been added to the fluxes of 26th and 27th January, respectively. Identified frequencies are marked with vertical dashed lines.}
    \label{fig:opt_spec}
\end{figure}

\subsection{Optical spectra}
Optical spectra at three epochs of observation are shown in Figure \ref{fig:opt_spec}. At all three epochs, the spectrum is dominated by Balmer emission lines from $H_\alpha$ to $H_\delta$. We have identified several lines of He I and Bowen fluorescence, as well as a couple of He II lines. The blend of the Na D1 and D2 lines exhibits an absorption feature. A single Gaussian was fitted to obtain the properties of the emission lines. The fluxes, FWHM, and equivalent width (EW) of the emission lines are given in Table \ref{tab:opt_spect}. The flux of He II 4686 $\AA$ varies during the course of three nights, and the FWHM steadily increases. The Balmer decrement (ratio of H$\beta$ to H$\alpha$ flux) values were 1.083, 1.048, and 1.002 on the three nights, respectively. The average ratio of HeII to H$\beta$ on three nights was $\sim$0.6.

\section{Discussion}\label{sec:dis}
A detailed timing analysis of V515 And has been performed using the high-cadence optical photometric data obtained from TESS. We have obtained the \psp and \pbe as 465.4721$\pm$0.0003 s and 488.6067$\pm$0.0004 s, respectively, which is in agreement with the earlier reported periods \citep{Bikmaev2006, Butters2008}. We obtained a \porb of 2.73116$\pm$0.00004 h, hence confirming its existence in the period gap. This makes V515 And the second confirmed IP  to lie in the period gap after the Paloma \citep{Schwarz2007, Joshi2016}.  The spin period of Paloma \citep{2023AJ....165...43L} is much longer than that of V515 And, whereas their orbital periods are very close to each other.

We explain the accretion mechanism based on the strength of \spin, \be, and their harmonics in the power spectra as discussed in \cite{Wynn1992, Ferrario1999}. They showed that the modulation on the \spin\ frequency is related to disc-fed accretion, whereas the  \be\ frequency presence is suggestive of stream-fed accretion. However, the presence of both \spin\ and \be\ frequencies is indicative of disc-overflow accretion \citep{Hellier1991, Hellier1993}. The \orb\ frequency in the power spectra can be attributed to the obscuration of the WD by the material rotating in the binary frame or the eclipse of the hotspot by the secondary \citep{Warner1986}.

The dominance of the power of \spin\ and \be\ frequencies was found to switch among different sectors, indicating a change between disc-fed and stream-fed dominance in the system V515 And. In sectors 17 and 57, \spin\ frequency was found to be dominant, suggesting disc-fed dominance, whereas in sector 84, the \be\ frequency became dominant, resulting in stream-fed dominance. The stream-fed dominance sustained up to the middle of sector 85, after which the accretion again changed to disc-fed dominance. This type of changing accretion geometry has also been observed in other IPs like V2400 Oph \citep{Joshi2019, Langford2022}, FO Aqr \citep{Hameury2017, Littlefield2020}, TX Col \citep{Rawat2021, Littlefield2021, Pandey2023}, V902 Mon \citep{Rawat2022}, V709 Cas \citep{Rao2026}, etc. The most likely reason for this change could be due to the variations in the mass accretion rate \citep{deMartino1995, Buckley1996, deMartino1999}. An enhanced mass accretion rate could cause more material to skim over the disc and directly hit the magnetosphere, resulting in stream-fed dominance, as we observe. When the mass accretion rate decreases, the amount of material accreted via stream becomes less in comparison to that accreted via disc, and we observe disc-fed dominance. Similar transitions in accretion geometry were also seen in the AAVSO-CV data by \cite{Covington2022}. This changing accretion geometry trend is also reflected in the spin pulse fraction, which is higher during periods of disc-fed dominance and lower during periods of stream-fed-dominated accretion.

\begin{figure}
    \centering
    \subfigure[ Burst duration versus peak luminosity]{\includegraphics[width=0.45\textwidth]{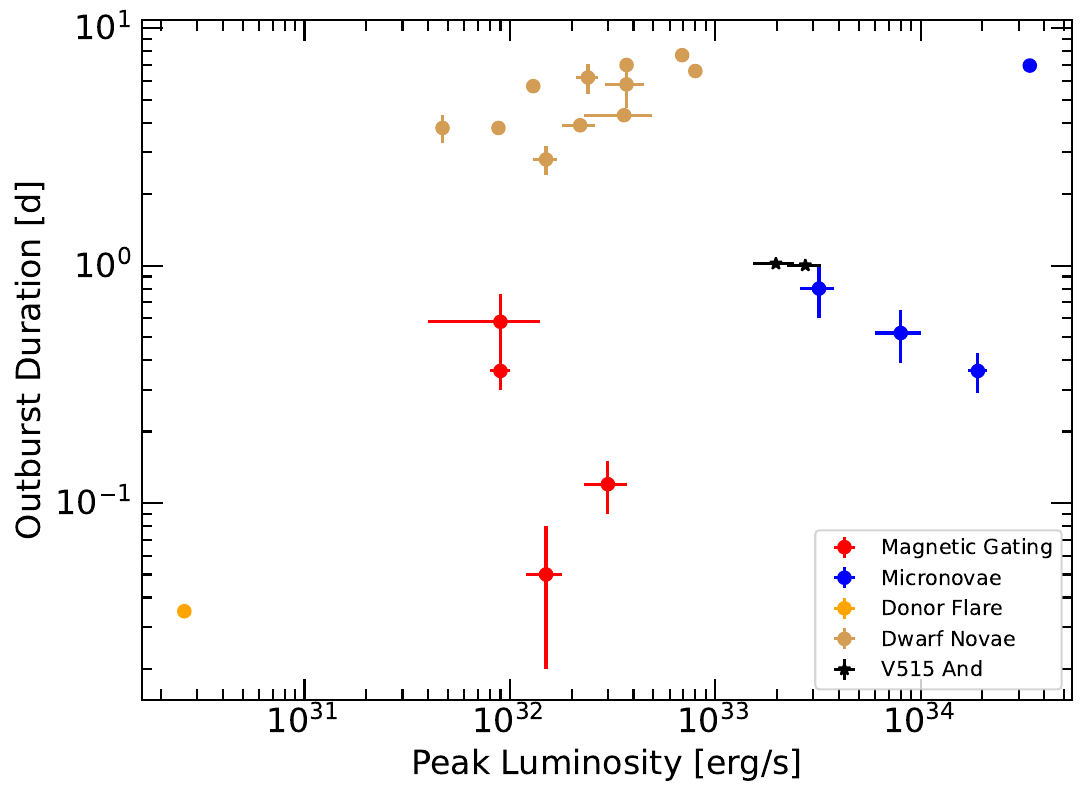}\label{}}
    \subfigure[Burst duration versus total energy ]{\includegraphics[width=0.45\textwidth]{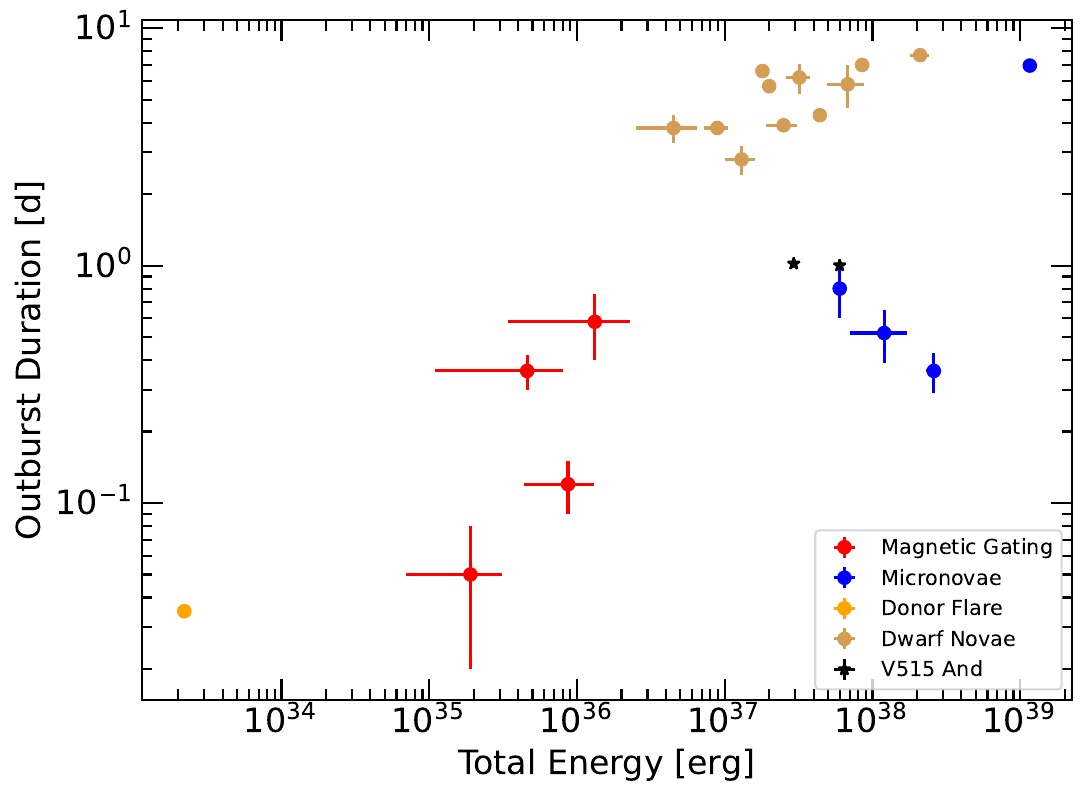}\label{}}
    \subfigure[Peak luminosity versus total energy]{\includegraphics[width=0.45\textwidth]{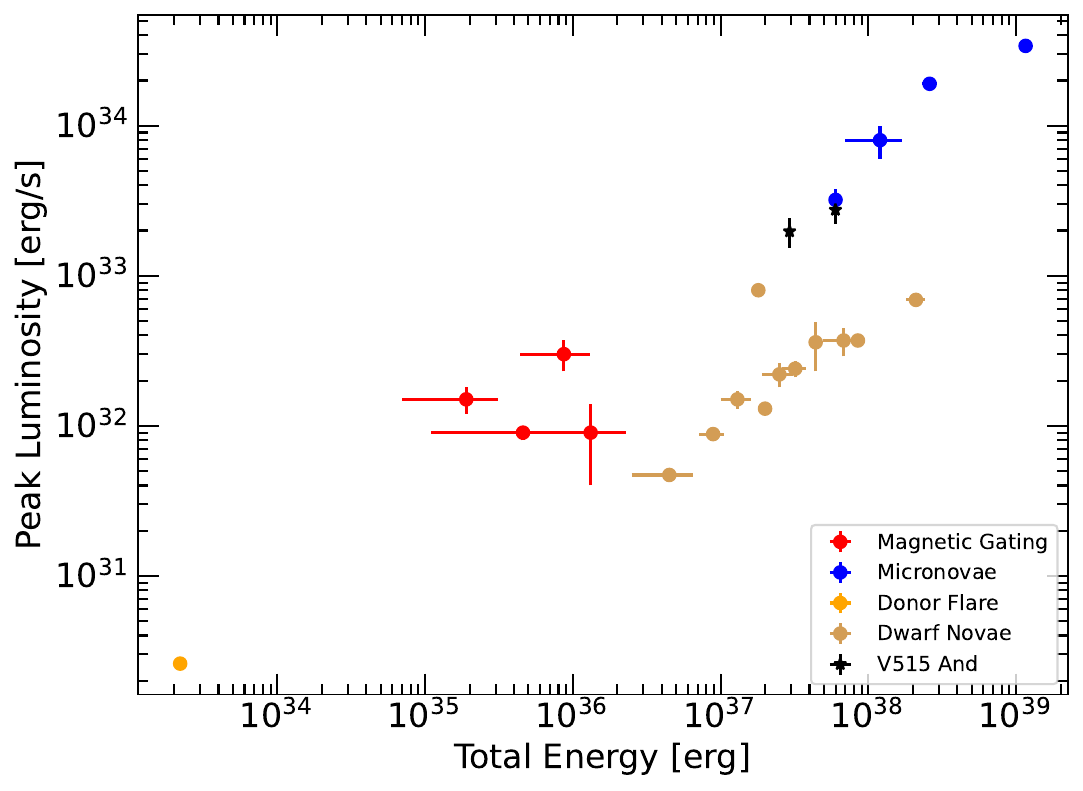}\label{}}
    \caption{Burst properties of CVs as adapted from Figure 2 of \protect\cite{Ilkiewicz2024}. The outburst properties of V515 And are shown in black markers.}
    \label{fig:burst_char_V515}
\end{figure}

Upon investigating the two outbursts observed in sector 57, we determined their total durations, peak optical luminosities, and total energies.  An attempt is made to classify these events into different types of outbursts.  Figure \ref{fig:burst_char_V515} presents the burst duration, peak optical luminosity, and total optical energy of both outbursts, plotted on the diagnostic diagrams from \cite{Ilkiewicz2024}. In all three panels, the parameters of V515 And overlap with the region corresponding to recently proposed micronova candidates. Micronovae are hypothesised to be localised thermonuclear events, resulting from the unstable nuclear burning of hydrogen that has accumulated on the poles of the WD \citep{Scaringi2022b}. These bursts have also been observed in other CVs such as TV Col, EI UMa, ASASSN-19bh \citep{Scaringi2022b}, CP Pup \citep{Veresvarska2024}, PBC J0801.2–4625 \citep{Irving2024}, DW Cnc \citep{Veresvarska2025}, in which  TV Col, EI UMa, and DW Cnc are IPs,  and ASASSN-19bh and PBC J0801.2–4625 are also highly likely to be IPs. This supports the necessary requirement of magnetic confinement of accreting material onto small regions \citep{Scaringi2022a}. Systems such as TV Col (\porb = 5.4864 h; \citealt{Hellier1993a}), EI Uma (\porb = 6.4344 h; \citealt{Thorstensen1986}), and PBC J0801.2-4625 (\porb = 5.906; \citealt{Irving2024}) lie above the period gap and while CP Pup (\porb = 1.47 h; \citealt{Orio2009}), and DW Cnc (\porb = 1.435 h; \citealt{Rodriguez2004}) lie below the period gap. For systems above the period gap, the dominant angular momentum loss (AML) mechanism is supposed to be magnetic braking (MB), which leads to high mass transfer rates ($\Dot{M}$). The model by \cite{Scaringi2022a} suggests that high mass accretion rates ($\Dot{m}$) might be favourable for triggering the micronova events.

For the systems below the period gap, the primary AML mechanism is gravitational radiation (GR). GR is not as efficient a mechanism as MB. Hence, the mass accretion rates are relatively lower. The typical mass accretion rate for IPs is of the order of  10$^{-9}$ M$_\odot$ yr$^{-1}$ \citep{Suleimanov2019}. \cite{Veresvarska2024, Veresvarska2025} obtained the mass accretion rate of CP Pup and DW Cnc in the order of 10$^{-10}$ and 10$^{-11}$ M$_\odot$ yr$^{-1}$, respectively. As V515 And lies in the period gap, it is expected to have an even lesser mass accretion rate; however, \cite{Suleimanov2019} estimated the mass accretion rate of V515 And as $2\times10^{-9}$ M$_\odot$ yr$^{-1}$. A high mass accretion rate is necessary to build up a critical mass of hydrogen ($M_{crit}$) that can trigger the burst. It is necessary to note that the mass transfer rate and mass accretion rate may not be the same, as some amount of material may undergo lateral spreading at the base of the magnetically confined poles or be accreted outside the magnetically confined poles or both \citep{Scaringi2022a}. Based on the formulations in \cite{Scaringi2022a}, despite having a lower mass accretion rate, an outburst may still be possible in V515 And if the magnetic confinement is confined to a very small region.

Optical spectra show hydrogen Balmer lines, along with He I and He II lines, in the emission.  The He II 4686 line is typically emitted from the inner accretion disc very close to the WD, or in the vicinity of the accretion column/curtain in the case of MCVs. The varying flux and FWHM trends suggest that there could be instabilities in this region, where the emitting region is becoming less energetic or smaller, or vice versa, while the gas velocities increase as it spirals closer to the WD. During the three nights, the invert Balmer decrement is found, indicating a high density and high optical depth accretion disc. The average ratio of HeII to H$\beta$ was greater than 0.4, indicating a magnetic nature of V515 And \citep{Silber1992}.

\section{Conclusions}\label{sec:conc}
The long-term photometric TESS and ground-based optical spectroscopic observations of V515 And lead us to the following conclusions:
\begin{itemize}[itemsep=\parskip]
    \item Using extensive TESS observations, we derive the orbital, spin and beat period of V515 And to be 2.73116$\pm$0.00004 h, 465.4721$\pm$0.0003 s, and 488.6067$\pm$0.0004 s, respectively. The detection of the orbital period of V515 And confirms its location in the period gap of \psp--\pbe plane.   
    \item The system displays a disc-overflow accretion geometry, with the relative dominance of disc-fed and stream-fed accretion varying among sectors and within individual sectors. 
    \item Two outburst events are detected during the TESS observations with burst durations of approximately one day. Given their peak optical luminosities on the order of 10$^{33}$ erg s$^{-1}$, total energies around 10$^{37}$ erg, and burst duration of $\sim$ 1 day, both events can be classified as micronova.
    \item Instabilities in the inner accretion disc are evident from varying flux levels of the He II 4686 line, and increasing FWHM points to a shrinking but faster-moving emitting region. 
\end{itemize}

\section*{Acknowledgements}
This paper includes data collected by the TESS mission funded by NASA's Science Mission Directorate. The study utilizes data from the 3.6 m Devasthal Optical Telescope (DOT), a National Facility, operated and overseen by the Aryabhatta Research Institute of Observational Sciences (ARIES), an autonomous Institute under the Department of Science and Technology, Government of India. Gratitude is extended to the scientific and technical personnel at ARIES DOT for their invaluable assistance. AJ acknowledges support from the Centro de Astrofisica y Tecnologias Afines (CATA) fellowship via grant Agencia Nacional de Investigacion Desarrollo (ANID), BASAL FB2100. We acknowledge the referee for reading our manuscript.

\section*{Data Availability}
The TESS data sets are publicly available in the TESS data archive at \url{https://archive.stsci.edu/missions-and-data/tess}. The optical spectroscopic data underlying this article will be shared on reasonable request to the corresponding author.



\bibliographystyle{mnras}
\bibliography{reference} 

@ARTICLE{Aizu1973,
       author = {{Aizu}, K.},
        title = "{X-Ray Emission Region of a White Dwarf with Accretion}",
      journal = {Progress of Theoretical Physics},
         year = 1973,
        month = apr,
       volume = {49},
       number = {4},
        pages = {1184-1194},
          doi = {10.1143/PTP.49.1184},
       adsurl = {https://ui.adsabs.harvard.edu/abs/1973PThPh..49.1184A}
}

@ARTICLE{Bikmaev2006,
       author = {{Bikmaev}, I.~F. and {Revnivtsev}, M.~G. and {Burenin}, R.~A. and {Sunyaev}, R.~A.},
        title = "{XSS J00564+4548 and IGR J00234+6141: New cataclysmic variables from the RXTE and INTEGRAL all-sky surveys}",
      journal = {Astronomy Letters},
         year = 2006,
        month = sep,
       volume = {32},
       number = {9},
        pages = {588-593},
          doi = {10.1134/S1063773706090039},
archivePrefix = {arXiv},
       eprint = {astro-ph/0603715},
 primaryClass = {astro-ph},
       adsurl = {https://ui.adsabs.harvard.edu/abs/2006AstL...32..588B}
}

@ARTICLE{Bonnet-Bidaud2009,
       author = {{Bonnet-Bidaud}, J.~M. and {de Martino}, D. and {Mouchet}, M.},
        title = "{XSS J0056+4548 : a hard X-ray intermediate polar in the period gap}",
      journal = {The Astronomer's Telegram},
         year = 2009,
        month = jan,
       volume = {1895},
        pages = {1},
       adsurl = {https://ui.adsabs.harvard.edu/abs/2009ATel.1895....1B}
}

@ARTICLE{Bruch2025,
       author = {{Bruch}, Albert},
        title = "{TESS Light Curves of Cataclysmic Variables. VI. Intermediate Polars}",
      journal = {\apjs},
         year = 2025,
        month = aug,
       volume = {279},
       number = {2},
          eid = {48},
        pages = {48},
          doi = {10.3847/1538-4365/addf41},
       adsurl = {https://ui.adsabs.harvard.edu/abs/2025ApJS..279...48B}
}

@INPROCEEDINGS{Buckley1996,
       author = {{Buckley}, D.~A.~H.},
        title = "{On the long-term light curve behaviour of the intermediate polar TX Col}",
    booktitle = {IAU Colloq. 158: Cataclysmic Variables and Related Objects},
         year = 1996,
       editor = {{Evans}, A. and {Wood}, Janet H.},
       series = {Astrophysics and Space Science Library},
       volume = {208},
        month = jan,
        pages = {185},
          doi = {10.1007/978-94-009-0325-8_55},
       adsurl = {https://ui.adsabs.harvard.edu/abs/1996ASSL..208..185B}
}

@ARTICLE{Butters2008,
       author = {{Butters}, O.~W. and {Norton}, A.~J. and {Hakala}, P. and {Mukai}, K. and {Barlow}, E.~J.},
        title = "{RXTE determination of the intermediate polar status of XSS J00564+4548, IGR J17195-4100, and XSS J12270-4859}",
      journal = {\aap},
         year = 2008,
        month = aug,
       volume = {487},
       number = {1},
        pages = {271-276},
          doi = {10.1051/0004-6361:200809942},
archivePrefix = {arXiv},
       eprint = {0806.0751},
 primaryClass = {astro-ph},
       adsurl = {https://ui.adsabs.harvard.edu/abs/2008A&A...487..271B}
}

@ARTICLE{Covington2022,
       author = {{Covington}, Ava E. and {Shaw}, Aarran W. and {Mukai}, Koji and {Littlefield}, Colin and {Heinke}, Craig O. and {Plotkin}, Richard M. and {Barrett}, Doug and {Boardman}, James and {Boyd}, David and {Brincat}, Stephen M. and {Carstens}, Rolf and {Collins}, Donald F. and {Cook}, Lewis M. and {Cooney}, Walter R. and {Fern{\'a}ndez}, David Cejudo and {Dufoer}, Sjoerd and {Dvorak}, Shawn and {Galdies}, Charles and {Goff}, William and {Hambsch}, Franz-Josef and {Johnston}, Steve and {Jones}, Jim and {Menzies}, Kenneth and {Monard}, Libert A.~G. and {Morelle}, Etienne and {Nelson}, Peter and {{\"O}gmen}, Yenal and {Rock}, John W. and {Sabo}, Richard and {Seargeant}, Jim and {Stone}, Geoffrey and {Ulowetz}, Joseph and {Vanmunster}, Tonny},
        title = "{Investigating the Low-flux States in Six Intermediate Polars}",
      journal = {\apj},
         year = 2022,
        month = apr,
       volume = {928},
       number = {2},
          eid = {164},
        pages = {164},
          doi = {10.3847/1538-4357/ac5682},
archivePrefix = {arXiv},
       eprint = {2202.08365},
 primaryClass = {astro-ph.SR},
       adsurl = {https://ui.adsabs.harvard.edu/abs/2022ApJ...928..164C}
}

@ARTICLE{Cropper1999,
       author = {{Cropper}, Mark and {Wu}, Kinwah and {Ramsay}, Gavin and {Kocabiyik}, Aysegul},
        title = "{Effects of gravity on the structure of post-shock accretion flows in magnetic cataclysmic variables}",
      journal = {\mnras},
         year = 1999,
        month = jul,
       volume = {306},
       number = {3},
        pages = {684-690},
          doi = {10.1046/j.1365-8711.1999.02570.x},
archivePrefix = {arXiv},
       eprint = {astro-ph/9902355},
 primaryClass = {astro-ph},
       adsurl = {https://ui.adsabs.harvard.edu/abs/1999MNRAS.306..684C}
}

@ARTICLE{deMartino1995,
       author = {{de Martino}, D. and {Buckely}, D.~A.~H. and {Mouchet}, M. and {Mukai}, K.},
        title = "{UV observations of the polar system RE 1938-461.}",
      journal = {\aap},
         year = 1995,
        month = jun,
       volume = {298},
        pages = {L5},
       adsurl = {https://ui.adsabs.harvard.edu/abs/1995A&A...298L...5D}
}

@ARTICLE{deMartino1999,
       author = {{de Martino}, D. and {Silvotti}, R. and {Buckley}, D.~A.~H. and {G{\"a}nsicke}, B.~T. and {Mouchet}, M. and {Mukai}, K. and {Rosen}, S.~R.},
        title = "{Time-resolved HST and IUE UV spectroscopy of the intermediate polar FO AQR}",
      journal = {\aap},
         year = 1999,
        month = oct,
       volume = {350},
        pages = {517-528},
          doi = {10.48550/arXiv.astro-ph/9909069},
archivePrefix = {arXiv},
       eprint = {astro-ph/9909069},
 primaryClass = {astro-ph},
       adsurl = {https://ui.adsabs.harvard.edu/abs/1999A&A...350..517D}
}

@ARTICLE{Ferrario1999,
       author = {{Ferrario}, Lilia and {Wickramasinghe}, D.~T.},
        title = "{The power of intermediate polars}",
      journal = {\mnras},
         year = 1999,
        month = oct,
       volume = {309},
       number = {2},
        pages = {517-527},
          doi = {10.1046/j.1365-8711.1999.02860.x},
       adsurl = {https://ui.adsabs.harvard.edu/abs/1999MNRAS.309..517F}
}

@ARTICLE{Gaia2023,
       author = {{Gaia Collaboration} and {Vallenari}, A. and {Brown}, A.~G.~A. and {Prusti}, T. and {de Bruijne}, J.~H.~J. and {Arenou}, F. and {Babusiaux}, C. and {Biermann}, M. and {Creevey}, O.~L. and {Ducourant}, C. and {Evans}, D.~W. and {Eyer}, L. and {Guerra}, R. and {Hutton}, A. and {Jordi}, C. and {Klioner}, S.~A. and {Lammers}, U.~L. and {Lindegren}, L. and {Luri}, X. and {Mignard}, F. and {Panem}, C. and {Pourbaix}, D. and {Randich}, S. and {Sartoretti}, P. and {Soubiran}, C. and {Tanga}, P. and {Walton}, N.~A. and {Bailer-Jones}, C.~A.~L. and {Bastian}, U. and {Drimmel}, R. and {Jansen}, F. and {Katz}, D. and {Lattanzi}, M.~G. and {van Leeuwen}, F. and {Bakker}, J. and {Cacciari}, C. and {Casta{\~n}eda}, J. and {De Angeli}, F. and {Fabricius}, C. and {Fouesneau}, M. and {Fr{\'e}mat}, Y. and {Galluccio}, L. and {Guerrier}, A. and {Heiter}, U. and {Masana}, E. and {Messineo}, R. and {Mowlavi}, N. and {Nicolas}, C. and {Nienartowicz}, K. and {Pailler}, F. and {Panuzzo}, P. and {Riclet}, F. and {Roux}, W. and {Seabroke}, G.~M. and {Sordo}, R. and {Th{\'e}venin}, F. and {Gracia-Abril}, G. and {Portell}, J. and {Teyssier}, D. and {Altmann}, M. and {Andrae}, R. and {Audard}, M. and {Bellas-Velidis}, I. and {Benson}, K. and {Berthier}, J. and {Blomme}, R. and {Burgess}, P.~W. and {Busonero}, D. and {Busso}, G. and {C{\'a}novas}, H. and {Carry}, B. and {Cellino}, A. and {Cheek}, N. and {Clementini}, G. and {Damerdji}, Y. and {Davidson}, M. and {de Teodoro}, P. and {Nu{\~n}ez Campos}, M. and {Delchambre}, L. and {Dell'Oro}, A. and {Esquej}, P. and {Fern{\'a}ndez-Hern{\'a}ndez}, J. and {Fraile}, E. and {Garabato}, D. and {Garc{\'\i}a-Lario}, P. and {Gosset}, E. and {Haigron}, R. and {Halbwachs}, J. -L. and {Hambly}, N.~C. and {Harrison}, D.~L. and {Hern{\'a}ndez}, J. and {Hestroffer}, D. and {Hodgkin}, S.~T. and {Holl}, B. and {Jan{\ss}en}, K. and {Jevardat de Fombelle}, G. and {Jordan}, S. and {Krone-Martins}, A. and {Lanzafame}, A.~C. and {L{\"o}ffler}, W. and {Marchal}, O. and {Marrese}, P.~M. and {Moitinho}, A. and {Muinonen}, K. and {Osborne}, P. and {Pancino}, E. and {Pauwels}, T. and {Recio-Blanco}, A. and {Reyl{\'e}}, C. and {Riello}, M. and {Rimoldini}, L. and {Roegiers}, T. and {Rybizki}, J. and {Sarro}, L.~M. and {Siopis}, C. and {Smith}, M. and {Sozzetti}, A. and {Utrilla}, E. and {van Leeuwen}, M. and {Abbas}, U. and {{\'A}brah{\'a}m}, P. and {Abreu Aramburu}, A. and {Aerts}, C. and {Aguado}, J.~J. and {Ajaj}, M. and {Aldea-Montero}, F. and {Altavilla}, G. and {{\'A}lvarez}, M.~A. and {Alves}, J. and {Anders}, F. and {Anderson}, R.~I. and {Anglada Varela}, E. and {Antoja}, T. and {Baines}, D. and {Baker}, S.~G. and {Balaguer-N{\'u}{\~n}ez}, L. and {Balbinot}, E. and {Balog}, Z. and {Barache}, C. and {Barbato}, D. and {Barros}, M. and {Barstow}, M.~A. and {Bartolom{\'e}}, S. and {Bassilana}, J. -L. and {Bauchet}, N. and {Becciani}, U. and {Bellazzini}, M. and {Berihuete}, A. and {Bernet}, M. and {Bertone}, S. and {Bianchi}, L. and {Binnenfeld}, A. and {Blanco-Cuaresma}, S. and {Blazere}, A. and {Boch}, T. and {Bombrun}, A. and {Bossini}, D. and {Bouquillon}, S. and {Bragaglia}, A. and {Bramante}, L. and {Breedt}, E. and {Bressan}, A. and {Brouillet}, N. and {Brugaletta}, E. and {Bucciarelli}, B. and {Burlacu}, A. and {Butkevich}, A.~G. and {Buzzi}, R. and {Caffau}, E. and {Cancelliere}, R. and {Cantat-Gaudin}, T. and {Carballo}, R. and {Carlucci}, T. and {Carnerero}, M.~I. and {Carrasco}, J.~M. and {Casamiquela}, L. and {Castellani}, M. and {Castro-Ginard}, A. and {Chaoul}, L. and {Charlot}, P. and {Chemin}, L. and {Chiaramida}, V. and {Chiavassa}, A. and {Chornay}, N. and {Comoretto}, G. and {Contursi}, G. and {Cooper}, W.~J. and {Cornez}, T. and {Cowell}, S. and {Crifo}, F. and {Cropper}, M. and {Crosta}, M. and {Crowley}, C. and {Dafonte}, C. and {Dapergolas}, A. and {David}, M. and {David}, P. and {de Laverny}, P. and {De Luise}, F. and {De March}, R.},
        title = "{Gaia Data Release 3. Summary of the content and survey properties}",
      journal = {\aap},
         year = 2023,
        month = jun,
       volume = {674},
          eid = {A1},
        pages = {A1},
          doi = {10.1051/0004-6361/202243940},
archivePrefix = {arXiv},
       eprint = {2208.00211},
 primaryClass = {astro-ph.GA},
       adsurl = {https://ui.adsabs.harvard.edu/abs/2023A&A...674A...1G}
}

@ARTICLE{Hameury2017,
       author = {{Hameury}, J. -M. and {Lasota}, J. -P.},
        title = "{The disappearance and reformation of the accretion disc during a low state of FO Aquarii}",
      journal = {\aap},
         year = 2017,
        month = oct,
       volume = {606},
          eid = {A7},
        pages = {A7},
          doi = {10.1051/0004-6361/201731226},
archivePrefix = {arXiv},
       eprint = {1707.00540},
 primaryClass = {astro-ph.SR},
       adsurl = {https://ui.adsabs.harvard.edu/abs/2017A&A...606A...7H}
}

@ARTICLE{Hellier1989a,
       author = {{Hellier}, Coel and {Mason}, Keith O. and {Cropper}, Mark},
        title = "{An eclipse in FO Aquarii.}",
      journal = {\mnras},
         year = 1989,
        month = mar,
       volume = {237},
        pages = {39P-44},
          doi = {10.1093/mnras/237.1.39P},
       adsurl = {https://ui.adsabs.harvard.edu/abs/1989MNRAS.237P..39H}
}

@ARTICLE{Hellier1989b,
       author = {{Hellier}, C. and {Mason}, K.~O. and {Smale}, A.~P. and {Corbet}, R.~H.~D. and {O'Donoghue}, D. and {Barrett}, P.~E. and {Warner}, B.},
        title = "{EX Hydrae in outburst.}",
      journal = {\mnras},
         year = 1989,
        month = jun,
       volume = {238},
        pages = {1107-1119},
          doi = {10.1093/mnras/238.4.1107},
       adsurl = {https://ui.adsabs.harvard.edu/abs/1989MNRAS.238.1107H}
}

@ARTICLE{Hellier1991,
       author = {{Hellier}, Coel},
        title = "{Do observations reveal accretion discs in intermediate polars ?}",
      journal = {\mnras},
         year = 1991,
        month = aug,
       volume = {251},
        pages = {693},
          doi = {10.1093/mnras/251.4.693},
       adsurl = {https://ui.adsabs.harvard.edu/abs/1991MNRAS.251..693H}
}

@ARTICLE{Hellier1993,
       author = {{Hellier}, Coel},
        title = "{Disc-overflow accretion in the intermediate polar FO Aquarii.}",
      journal = {\mnras},
         year = 1993,
        month = dec,
       volume = {265},
        pages = {L35-L39},
          doi = {10.1093/mnras/265.1.L35},
       adsurl = {https://ui.adsabs.harvard.edu/abs/1993MNRAS.265L..35H}
}

@ARTICLE{Hellier1993a,
       author = {{Hellier}, C.},
        title = "{The four periodicities of the cataclysmic variable TV Columbae.}",
      journal = {\mnras},
         year = 1993,
        month = sep,
       volume = {264},
        pages = {132-144},
          doi = {10.1093/mnras/264.1.132},
       adsurl = {https://ui.adsabs.harvard.edu/abs/1993MNRAS.264..132H}
}

@ARTICLE{Ilkiewicz2024,
       author = {{I{\l}kiewicz}, Krystian and {Scaringi}, Simone and {Veresvarska}, Martina and {De Martino}, Domitilla and {Littlefield}, Colin and {Knigge}, Christian and {Paice}, John A. and {Sahu}, Anwesha},
        title = "{Classifying Optical (Out)bursts in Cataclysmic Variables: The Distinct Observational Characteristics of Dwarf Novae, Micronovae, Stellar Flares, and Magnetic Gating}",
      journal = {\apjl},
         year = 2024,
        month = feb,
       volume = {962},
       number = {2},
          eid = {L34},
        pages = {L34},
          doi = {10.3847/2041-8213/ad243c},
archivePrefix = {arXiv},
       eprint = {2402.00553},
 primaryClass = {astro-ph.SR},
       adsurl = {https://ui.adsabs.harvard.edu/abs/2024ApJ...962L..34I}
}

@ARTICLE{Irving2024,
       author = {{Irving}, Z.~A. and {Altamirano}, D. and {Scaringi}, S. and {Veresvarska}, M. and {Knigge}, C. and {Castro Segura}, N. and {De Martino}, D. and {I{\l}kiewicz}, K.},
        title = "{Burst-induced spin variations in the accreting magnetic white dwarf PBC J0801.2-4625}",
      journal = {\mnras},
         year = 2024,
        month = jun,
       volume = {530},
       number = {4},
        pages = {3974-3985},
          doi = {10.1093/mnras/stae1103},
       adsurl = {https://ui.adsabs.harvard.edu/abs/2024MNRAS.530.3974I}
}

@INPROCEEDINGS{Jenkins2016,
       author = {{Jenkins}, Jon M. and {Twicken}, Joseph D. and {McCauliff}, Sean and {Campbell}, Jennifer and {Sanderfer}, Dwight and {Lung}, David and {Mansouri-Samani}, Masoud and {Girouard}, Forrest and {Tenenbaum}, Peter and {Klaus}, Todd and {Smith}, Jeffrey C. and {Caldwell}, Douglas A. and {Chacon}, A.~D. and {Henze}, Christopher and {Heiges}, Cory and {Latham}, David W. and {Morgan}, Edward and {Swade}, Daryl and {Rinehart}, Stephen and {Vanderspek}, Roland},
        title = "{The TESS science processing operations center}",
    booktitle = {Software and Cyberinfrastructure for Astronomy IV},
         year = 2016,
       editor = {{Chiozzi}, Gianluca and {Guzman}, Juan C.},
       series = {Society of Photo-Optical Instrumentation Engineers (SPIE) Conference Series},
       volume = {9913},
        month = aug,
          eid = {99133E},
        pages = {99133E},
          doi = {10.1117/12.2233418},
       adsurl = {https://ui.adsabs.harvard.edu/abs/2016SPIE.9913E..3EJ}
}

@ARTICLE{Joshi2016,
       author = {{Joshi}, Arti and {Pandey}, J.~C. and {Singh}, K.~P. and {Agrawal}, P.~C.},
        title = "{PALOMA: A Magnetic CV between Polars and Intermediate Polars}",
      journal = {\apj},
         year = 2016,
        month = oct,
       volume = {830},
       number = {2},
          eid = {56},
        pages = {56},
          doi = {10.3847/0004-637X/830/2/56},
archivePrefix = {arXiv},
       eprint = {1610.00557},
 primaryClass = {astro-ph.GA},
       adsurl = {https://ui.adsabs.harvard.edu/abs/2016ApJ...830...56J}
}

@ARTICLE{Joshi2019,
       author = {{Joshi}, Arti and {Pandey}, J.~C. and {Singh}, Harinder P.},
        title = "{X-Ray Observations of an Intermediate Polar V2400 Oph}",
      journal = {AJ},
         year = 2019,
        month = jul,
       volume = {158},
       number = {1},
          eid = {11},
        pages = {11},
          doi = {10.3847/1538-3881/ab1ea6},
       adsurl = {https://ui.adsabs.harvard.edu/abs/2019AJ....158...11J}
}

@ARTICLE{Kalman2025,
       author = {{K{\'a}lm{\'a}n}, Szil{\'a}rd and {Csizmadia}, Szil{\'a}rd and {P{\'a}l}, Andr{\'a}s and {Szab{\'o}}, Gyula M.},
        title = "{Detection of a Peculiar Noise Type in the TESS ``Fast'' Light Curves}",
      journal = {Research Notes of the American Astronomical Society},
         year = 2025,
        month = feb,
       volume = {9},
       number = {2},
          eid = {33},
        pages = {33},
          doi = {10.3847/2515-5172/adb3aa},
archivePrefix = {arXiv},
       eprint = {2502.10326},
 primaryClass = {astro-ph.EP},
       adsurl = {https://ui.adsabs.harvard.edu/abs/2025RNAAS...9...33K}
}

@ARTICLE{Kolb1993,
       author = {{Kolb}, U.},
        title = "{A model for the intrinsic population of cataclysmic variables}",
      journal = {\aap},
         year = 1993,
        month = apr,
       volume = {271},
        pages = {149},
       adsurl = {https://ui.adsabs.harvard.edu/abs/1993A&A...271..149K}
}

@ARTICLE{Kozhevnikov2012,
       author = {{Kozhevnikov}, V.~P.},
        title = "{An extensive photometric study of the recently discovered intermediate polar V515 And (XSS J00564+4548)}",
      journal = {\mnras},
         year = 2012,
        month = may,
       volume = {422},
       number = {2},
        pages = {1518-1526},
          doi = {10.1111/j.1365-2966.2012.20725.x},
archivePrefix = {arXiv},
       eprint = {1202.2493},
 primaryClass = {astro-ph.SR},
       adsurl = {https://ui.adsabs.harvard.edu/abs/2012MNRAS.422.1518K}
}

@ARTICLE{Kumar2018,
       author = {{Kumar}, Brijesh and {Omar}, Amitesh and {Maheswar}, Gopinathan and {Pandey}, Anil Kumar and {Sagar}, Ram and {Uddin}, Wahab and {Sanwal}, Basant Ballabh and {Bangia}, Tarun and {Kumar}, Tripurari Satyanarayana and {Yadav}, Shobhit and {Sahu}, Sanjit and {Pant}, Jayshreekar and {Reddy}, Bheemireddy Krishna and {Gupta}, Alok Chandra and {Chand}, Hum and {Pandey}, Jeewan Chandra and {Joshi}, Mohit Kumar and {Jaiswar}, Mukeshkuma and {Nanjappa}, Nandish and {Purushottam} and {Yadav}, Rama Kant Singh and {Sharma}, Saurabh and {Pandey}, Shashi Bhushan and {Joshi}, Santosh and {Joshi}, Yogesh Chandra and {Lata}, Sneh and {Mehdi}, Biman Jyoti and {Misra}, Kuntal and {Singh}, Mahendra},
        title = "{3.6-m Devasthal Optical Telescope Project: Completion and first results}",
      journal = {Bulletin de la Societe Royale des Sciences de Liege},
         year = 2018,
        month = apr,
       volume = {87},
        pages = {29-41},
       adsurl = {https://ui.adsabs.harvard.edu/abs/2018BSRSL..87...29K}
}

@ARTICLE{Langford2022,
       author = {{Langford}, Andrew and {Littlefield}, Colin and {Garnavich}, Peter and {Kennedy}, Mark R. and {Scaringi}, Simone and {Szkody}, Paula},
        title = "{Searching for Diamagnetic Blob Accretion in the 74 day K2 Observation of V2400 Ophiuchi}",
      journal = {\aj},
         year = 2022,
        month = jan,
       volume = {163},
       number = {1},
          eid = {4},
        pages = {4},
          doi = {10.3847/1538-3881/ac3010},
archivePrefix = {arXiv},
       eprint = {2112.07729},
 primaryClass = {astro-ph.SR},
       adsurl = {https://ui.adsabs.harvard.edu/abs/2022AJ....163....4L}
}

@ARTICLE{Littlefield2020,
       author = {{Littlefield}, Colin and {Garnavich}, Peter and {Kennedy}, Mark R. and {Patterson}, Joseph and {Kemp}, Jonathan and {Stiller}, Robert A. and {Hambsch}, Franz-Josef and {Heras}, Te{\'o}filo Arranz and {Myers}, Gordon and {Stone}, Geoffrey and {Sj{\"o}berg}, George and {Dvorak}, Shawn and {Nelson}, Peter and {Popov}, Velimir and {Bonnardeau}, Michel and {Vanmunster}, Tonny and {de Miguel}, Enrique and {Alton}, Kevin B. and {Harris}, Barbara and {Cook}, Lewis M. and {Graham}, Keith A. and {Brincat}, Stephen M. and {Lane}, David J. and {Foster}, James and {Pickard}, Roger and {Sabo}, Richard and {Vietje}, Brad and {Lemay}, Damien and {Briol}, John and {Krumm}, Nathan and {Dadighat}, Michelle and {Goff}, William and {Solomon}, Rob and {Padovan}, Stefano and {Bolt}, Greg and {Kardasis}, Emmanuel and {Deback{\`e}re}, Andr{\'e} and {Thrush}, Jeff and {Stein}, William and {Walter}, Bradley and {Coulter}, Daniel and {Tsehmeystrenko}, Valery and {Gout}, Jean-Fran{\c{c}}ois and {Lewin}, Pablo and {Galdies}, Charles and {Fernandez}, David Cejudo and {Walker}, Gary and {Boardman}, James, Jr. and {Pellett}, Emil},
        title = "{The Rise and Fall of the King: The Correlation between FO Aquarii's Low States and the White Dwarf's Spin-down}",
      journal = {APJ},
         year = 2020,
        month = jun,
       volume = {896},
       number = {2},
          eid = {116},
        pages = {116},
          doi = {10.3847/1538-4357/ab9197},
archivePrefix = {arXiv},
       eprint = {1904.11505},
 primaryClass = {astro-ph.SR},
       adsurl = {https://ui.adsabs.harvard.edu/abs/2020ApJ...896..116L}
}

@ARTICLE{Littlefield2021,
       author = {{Littlefield}, Colin and {Scaringi}, Simone and {Garnavich}, Peter and {Szkody}, Paula and {Kennedy}, Mark R. and {I{\l}kiewicz}, Krystian and {Mason}, Paul A.},
        title = "{Quasi-periodic Oscillations in the TESS Light Curve of TX Col, a Diskless Intermediate Polar on the Precipice of Forming an Accretion Disk}",
      journal = {\aj},
         year = 2021,
        month = aug,
       volume = {162},
       number = {2},
          eid = {49},
        pages = {49},
          doi = {10.3847/1538-3881/ac062b},
archivePrefix = {arXiv},
       eprint = {2104.14140},
 primaryClass = {astro-ph.SR},
       adsurl = {https://ui.adsabs.harvard.edu/abs/2021AJ....162...49L}
}

@ARTICLE{Lomb1976,
       author = {{Lomb}, N.~R.},
        title = "{Least-Squares Frequency Analysis of Unequally Spaced Data}",
      journal = {\apss},
         year = 1976,
        month = feb,
       volume = {39},
       number = {2},
        pages = {447-462},
          doi = {10.1007/BF00648343},
       adsurl = {https://ui.adsabs.harvard.edu/abs/1976Ap&SS..39..447L}
}

@ARTICLE{Omar2019,
       author = {{Omar}, Amitesh and {Kumar}, T.~S. and {Krishna Reddy}, B. and {Pant}, Jayshreekar and {Mahto}, Manoj},
        title = "{First-light images from low dispersion spectrograph-cum-imager on 3.6-meter Devasthal Optical Telescope}",
      journal = {arXiv e-prints},
         year = 2019,
        month = feb,
          eid = {arXiv:1902.05857},
        pages = {arXiv:1902.05857},
          doi = {10.48550/arXiv.1902.05857},
archivePrefix = {arXiv},
       eprint = {1902.05857},
 primaryClass = {astro-ph.IM},
       adsurl = {https://ui.adsabs.harvard.edu/abs/2019arXiv190205857O}
}

@ARTICLE{Orio2009,
       author = {{Orio}, M. and {Mukai}, K. and {Bianchini}, A. and {de Martino}, D. and {Howell}, S.},
        title = "{New X-Ray Observations of the Old Nova CP Puppis and of the More Recent Nova V351 Puppis}",
      journal = {\apj},
         year = 2009,
        month = jan,
       volume = {690},
       number = {2},
        pages = {1753-1763},
          doi = {10.1088/0004-637X/690/2/1753},
archivePrefix = {arXiv},
       eprint = {0809.3992},
 primaryClass = {astro-ph},
       adsurl = {https://ui.adsabs.harvard.edu/abs/2009ApJ...690.1753O}
}

@ARTICLE{Panchal2023,
       author = {{Panchal}, Dimple and {Kumar}, Tripurari S. and {Omar}, Amitesh and {Misra}, Kuntal},
        title = "{Characterization of a deep-depletion 4K {\texttimes} 4K charge-coupled device detector system designed for ARIES Devasthal faint object spectrograph}",
      journal = {Journal of Astronomical Telescopes, Instruments, and Systems},
         year = 2023,
        month = jan,
       volume = {9},
          eid = {018002},
        pages = {018002},
          doi = {10.1117/1.JATIS.9.1.018002},
archivePrefix = {arXiv},
       eprint = {2301.08746},
 primaryClass = {astro-ph.IM},
       adsurl = {https://ui.adsabs.harvard.edu/abs/2023JATIS...9a8002P}
}

@ARTICLE{Pandey2023,
       author = {{Pandey}, Jeewan Chandra and {Rawat}, Nikita and {Rao}, Srinivas M. and {Joshi}, Arti and {Singh}, Sadhana},
        title = "{X-ray Observations of the Intermediate Polar TX Col}",
      journal = {Bulletin de la Societe Royale des Sciences de Liege},
         year = 2024,
        month = jun,
       volume = {93},
       number = {2},
        pages = {243-249},
          doi = {10.25518/0037-9565.11662},
archivePrefix = {arXiv},
       eprint = {2309.03674},
 primaryClass = {astro-ph.HE},
       adsurl = {https://ui.adsabs.harvard.edu/abs/2024BSRSL..93..243P}
}

@ARTICLE{Patterson1994,
       author = {{Patterson}, Joseph},
        title = "{The DQ Herculis Stars}",
      journal = {\pasp},
         year = 1994,
        month = mar,
       volume = {106},
        pages = {209},
          doi = {10.1086/133375},
       adsurl = {https://ui.adsabs.harvard.edu/abs/1994PASP..106..209P}
}

@ARTICLE{Rao2026,
       author = {{Rao}, Srinivas M. and {Pandey}, Jeewan C. and {Rawat}, Nikita and {Joshi}, Arti and {Singh}, Ajay Kumar},
        title = "{Long-term optical photometry of V709 Cas using TESS: Refined periods and accretion geometry}",
      journal = {\na},
         year = 2026,
        month = jan,
       volume = {122},
          eid = {102481},
        pages = {102481},
          doi = {10.1016/j.newast.2025.102481},
archivePrefix = {arXiv},
       eprint = {2507.19441},
 primaryClass = {astro-ph.SR},
       adsurl = {https://ui.adsabs.harvard.edu/abs/2026NewA..12202481R}
}

@ARTICLE{Rawat2021,
       author = {{Rawat}, Nikita and {Pandey}, J.~C. and {Joshi}, Arti},
        title = "{TESS Observations of TX Col: Rapidly Varying Accretion Flow}",
      journal = {APJ},
         year = 2021,
        month = may,
       volume = {912},
       number = {1},
          eid = {78},
        pages = {78},
          doi = {10.3847/1538-4357/abedae},
archivePrefix = {arXiv},
       eprint = {2104.06944},
 primaryClass = {astro-ph.SR},
       adsurl = {https://ui.adsabs.harvard.edu/abs/2021ApJ...912...78R}
}

@ARTICLE{Rawat2022,
       author = {{Rawat}, Nikita and {Pandey}, J.~C. and {Joshi}, Arti and {Yadava}, Umesh},
        title = "{A step towards unveiling the nature of three cataclysmic variables: LS Cam, V902 Mon, and SWIFT J0746.3-1608}",
      journal = {MNRAS},
         year = 2022,
        month = jun,
       volume = {512},
       number = {4},
        pages = {6054-6066},
          doi = {10.1093/mnras/stac844},
archivePrefix = {arXiv},
       eprint = {2203.17088},
 primaryClass = {astro-ph.SR},
       adsurl = {https://ui.adsabs.harvard.edu/abs/2022MNRAS.512.6054R}
}

@ARTICLE{Ricker2015,
       author = {{Ricker}, George R. and {Winn}, Joshua N. and {Vanderspek}, Roland and {Latham}, David W. and {Bakos}, G{\'a}sp{\'a}r {\'A}. and {Bean}, Jacob L. and {Berta-Thompson}, Zachory K. and {Brown}, Timothy M. and {Buchhave}, Lars and {Butler}, Nathaniel R. and {Butler}, R. Paul and {Chaplin}, William J. and {Charbonneau}, David and {Christensen-Dalsgaard}, J{\o}rgen and {Clampin}, Mark and {Deming}, Drake and {Doty}, John and {De Lee}, Nathan and {Dressing}, Courtney and {Dunham}, Edward W. and {Endl}, Michael and {Fressin}, Francois and {Ge}, Jian and {Henning}, Thomas and {Holman}, Matthew J. and {Howard}, Andrew W. and {Ida}, Shigeru and {Jenkins}, Jon M. and {Jernigan}, Garrett and {Johnson}, John Asher and {Kaltenegger}, Lisa and {Kawai}, Nobuyuki and {Kjeldsen}, Hans and {Laughlin}, Gregory and {Levine}, Alan M. and {Lin}, Douglas and {Lissauer}, Jack J. and {MacQueen}, Phillip and {Marcy}, Geoffrey and {McCullough}, Peter R. and {Morton}, Timothy D. and {Narita}, Norio and {Paegert}, Martin and {Palle}, Enric and {Pepe}, Francesco and {Pepper}, Joshua and {Quirrenbach}, Andreas and {Rinehart}, Stephen A. and {Sasselov}, Dimitar and {Sato}, Bun'ei and {Seager}, Sara and {Sozzetti}, Alessandro and {Stassun}, Keivan G. and {Sullivan}, Peter and {Szentgyorgyi}, Andrew and {Torres}, Guillermo and {Udry}, Stephane and {Villasenor}, Joel},
        title = "{Transiting Exoplanet Survey Satellite (TESS)}",
      journal = {Journal of Astronomical Telescopes, Instruments, and Systems},
         year = 2015,
        month = jan,
       volume = {1},
          eid = {014003},
        pages = {014003},
          doi = {10.1117/1.JATIS.1.1.014003},
       adsurl = {https://ui.adsabs.harvard.edu/abs/2015JATIS...1a4003R}
}

@ARTICLE{Rodriguez2004,
       author = {{Rodr{\'\i}guez-Gil}, P. and {G{\"a}nsicke}, B.~T. and {Araujo-Betancor}, S. and {Casares}, J.},
        title = "{DW Cancri: a magnetic VY Scl star with an orbital period of 86 min}",
      journal = {\mnras},
         year = 2004,
        month = mar,
       volume = {349},
       number = {1},
        pages = {367-374},
          doi = {10.1111/j.1365-2966.2004.07512.x},
archivePrefix = {arXiv},
       eprint = {astro-ph/0312149},
 primaryClass = {astro-ph},
       adsurl = {https://ui.adsabs.harvard.edu/abs/2004MNRAS.349..367R}
}

@ARTICLE{Rosen1988,
       author = {{Rosen}, S.~R. and {Mason}, K.~O. and {Cordova}, F.~A.},
        title = "{EXOSAT X-ray observations of the eclipsing magnetic cataclysmic variable EX Hya.}",
      journal = {\mnras},
         year = 1988,
        month = mar,
       volume = {231},
        pages = {549-573},
          doi = {10.1093/mnras/231.3.549},
       adsurl = {https://ui.adsabs.harvard.edu/abs/1988MNRAS.231..549R}
}

@ARTICLE{Scargle1982,
       author = {{Scargle}, J.~D.},
        title = "{Studies in astronomical time series analysis. II. Statistical aspects of spectral analysis of unevenly spaced data.}",
      journal = {\apj},
         year = 1982,
        month = dec,
       volume = {263},
        pages = {835-853},
          doi = {10.1086/160554},
       adsurl = {https://ui.adsabs.harvard.edu/abs/1982ApJ...263..835S}
}

@ARTICLE{Scargle1998,
       author = {{Scargle}, Jeffrey D.},
        title = "{Studies in Astronomical Time Series Analysis. V. Bayesian Blocks, a New Method to Analyze Structure in Photon Counting Data}",
      journal = {\apj},
         year = 1998,
        month = sep,
       volume = {504},
       number = {1},
        pages = {405-418},
          doi = {10.1086/306064},
archivePrefix = {arXiv},
       eprint = {astro-ph/9711233},
 primaryClass = {astro-ph},
       adsurl = {https://ui.adsabs.harvard.edu/abs/1998ApJ...504..405S}
}

@ARTICLE{Scargle2013,
       author = {{Scargle}, Jeffrey D. and {Norris}, Jay P. and {Jackson}, Brad and {Chiang}, James},
        title = "{The Bayesian Block Algorithm}",
      journal = {arXiv e-prints},
         year = 2013,
        month = apr,
          eid = {arXiv:1304.2818},
        pages = {arXiv:1304.2818},
          doi = {10.48550/arXiv.1304.2818},
archivePrefix = {arXiv},
       eprint = {1304.2818},
 primaryClass = {astro-ph.IM},
       adsurl = {https://ui.adsabs.harvard.edu/abs/2013arXiv1304.2818S}
}

@ARTICLE{Scaringi2022a,
       author = {{Scaringi}, S. and {Groot}, P.~J. and {Knigge}, C. and {Lasota}, J. -P. and {de Martino}, D. and {Cavecchi}, Y. and {Buckley}, D.~A.~H. and {Camisassa}, M.~E.},
        title = "{Triggering micronovae through magnetically confined accretion flows in accreting white dwarfs}",
      journal = {\mnras},
         year = 2022,
        month = jul,
       volume = {514},
       number = {1},
        pages = {L11-L15},
          doi = {10.1093/mnrasl/slac042},
archivePrefix = {arXiv},
       eprint = {2204.09073},
 primaryClass = {astro-ph.HE},
       adsurl = {https://ui.adsabs.harvard.edu/abs/2022MNRAS.514L..11S}
}

@ARTICLE{Scaringi2022b,
       author = {{Scaringi}, S. and {Groot}, P.~J. and {Knigge}, C. and {Bird}, A.~J. and {Breedt}, E. and {Buckley}, D.~A.~H. and {Cavecchi}, Y. and {Degenaar}, N.~D. and {de Martino}, D. and {Done}, C. and {Fratta}, M. and {I{\l}kiewicz}, K. and {Koerding}, E. and {Lasota}, J. -P. and {Littlefield}, C. and {Manara}, C.~F. and {O'Brien}, M. and {Szkody}, P. and {Timmes}, F.~X.},
        title = "{Localized thermonuclear bursts from accreting magnetic white dwarfs}",
      journal = {\nat},
         year = 2022,
        month = apr,
       volume = {604},
       number = {7906},
        pages = {447-450},
          doi = {10.1038/s41586-022-04495-6},
archivePrefix = {arXiv},
       eprint = {2204.09070},
 primaryClass = {astro-ph.HE},
       adsurl = {https://ui.adsabs.harvard.edu/abs/2022Natur.604..447S}
}

@ARTICLE{Schreiber2024,
       author = {{Schreiber}, Matthias R. and {Belloni}, Diogo and {Schwope}, Axel D.},
        title = "{The cataclysmic variable orbital period gap: More evident than ever}",
      journal = {\aap},
         year = 2024,
        month = feb,
       volume = {682},
          eid = {L7},
        pages = {L7},
          doi = {10.1051/0004-6361/202348807},
archivePrefix = {arXiv},
       eprint = {2402.02076},
 primaryClass = {astro-ph.SR},
       adsurl = {https://ui.adsabs.harvard.edu/abs/2024A&A...682L...7S}
}

@ARTICLE{Schwarz2007,
       author = {{Schwarz}, R. and {Schwope}, A.~D. and {Staude}, A. and {Rau}, A. and {Hasinger}, G. and {Urrutia}, T. and {Motch}, C.},
        title = "{Paloma (RX J0524+42): the missing link in magnetic CV evolution?}",
      journal = {\aap},
         year = 2007,
        month = oct,
       volume = {473},
       number = {2},
        pages = {511-521},
          doi = {10.1051/0004-6361:20077684},
       adsurl = {https://ui.adsabs.harvard.edu/abs/2007A&A...473..511S}
}

@PHDTHESIS{Silber1992,
       author = {{Silber}, Andrew D.},
        title = "{Studies of an X-Ray Selected Sample of Cataclysmic Variables.}",
       school = {Massachusetts Institute of Technology},
         year = 1992,
        month = jan,
       adsurl = {https://ui.adsabs.harvard.edu/abs/1992PhDT.......119S}
}

@ARTICLE{Suleimanov2019,
       author = {{Suleimanov}, Valery F. and {Doroshenko}, Victor and {Werner}, Klaus},
        title = "{Hard X-ray view on intermediate polars in the Gaia era}",
      journal = {\mnras},
         year = 2019,
        month = jan,
       volume = {482},
       number = {3},
        pages = {3622-3635},
          doi = {10.1093/mnras/sty2952},
archivePrefix = {arXiv},
       eprint = {1809.05740},
 primaryClass = {astro-ph.HE},
       adsurl = {https://ui.adsabs.harvard.edu/abs/2019MNRAS.482.3622S}
}

@ARTICLE{Thorstensen1986,
       author = {{Thorstensen}, J.~R.},
        title = "{Orbital studies of cataclysmic binaries. II. Three objects from the Palomar-Green sample.}",
      journal = {\aj},
         year = 1986,
        month = apr,
       volume = {91},
        pages = {940-950},
          doi = {10.1086/114070},
       adsurl = {https://ui.adsabs.harvard.edu/abs/1986AJ.....91..940T}
}

@INPROCEEDINGS{Tody1986,
       author = {{Tody}, Doug},
        title = "{The IRAF Data Reduction and Analysis System}",
    booktitle = {Instrumentation in astronomy VI},
         year = 1986,
       editor = {{Crawford}, David L.},
       series = {Society of Photo-Optical Instrumentation Engineers (SPIE) Conference Series},
       volume = {627},
        month = jan,
        pages = {733},
          doi = {10.1117/12.968154},
       adsurl = {https://ui.adsabs.harvard.edu/abs/1986SPIE..627..733T}
}

@ARTICLE{Vaughan2005,
       author = {{Vaughan}, S.},
        title = "{A simple test for periodic signals in red noise}",
      journal = {\aap},
         year = 2005,
        month = feb,
       volume = {431},
        pages = {391-403},
          doi = {10.1051/0004-6361:20041453},
archivePrefix = {arXiv},
       eprint = {astro-ph/0412697},
 primaryClass = {astro-ph},
       adsurl = {https://ui.adsabs.harvard.edu/abs/2005A&A...431..391V}
}

\label{lastpage}
\end{document}